\documentclass[aps,pre,twocolumn,superscriptaddress]{revtex4-2}
\usepackage{tikz}
\usepackage[english]{babel}
\usepackage[utf8]{inputenc}
\usepackage[T1]{fontenc}
\usepackage{indentfirst}
\usepackage{amsmath}
\usepackage{amssymb}
\usepackage{amsthm}
\usepackage{proof}
\usepackage{eufrak}
\usepackage{graphicx}

\allowhyphens

\newcommand{\sech}{\text{sech}}
\newcommand{\La}{\mathcal{L}}
\newcommand{\Ga}{\mathcal{G}}
\newcommand{\Ra}{\mathcal{R}}
\newcommand{\Pe}{\mathcal{P}}
\newcommand{\UU}{\mathcal{U}}
\newcommand{\complex}{\mathbb{C}}
\newcommand{\egesz}{\mathbb{Z}}

\newcommand{\valos}{\mathbb{R}}

\newcommand{\ordo}{\mathcal{O}}

\newcommand{\sn}{\mathrm{sn}}
\newcommand{\cn}{\mathrm{cn}}
\newcommand{\dn}{\mathrm{dn}}

\newcommand{\ket}[1]{{\left|#1\right\rangle}}

\newtheorem{definition}{Definition}
\newtheorem{conjecture}{Conjecture}
\newtheorem{theorem}{Theorem}


\usepackage{ifpdf}

\ifpdf
\usepackage{epstopdf}
\usepackage[pdftex,colorlinks,urlcolor=blue,citecolor=blue,linkcolor=blue]{hyperref}
\else
\usepackage[hypertex,colorlinks,urlcolor=blue,citecolor=blue,linkcolor=blue]{hyperref}
\fi
\begin{document}
\numberwithin{equation}{section}

\title{Integrable spin chains and cellular automata with medium range interaction}

\author{Tam\'as Gombor}
\affiliation{MTA-ELTE “Momentum” Integrable Quantum Dynamics Research Group, Department of Theoretical Physics, Eötvös
  Loránd University}
\affiliation{Holographic QFT Group, Wigner Research Centre for Physics, Budapest, Hungary}
\author{Bal\'azs Pozsgay}
\affiliation{MTA-ELTE “Momentum” Integrable Quantum Dynamics Research Group, Department of Theoretical Physics, Eötvös Loránd University}

\begin{abstract}
We study integrable spin chains and quantum and classical cellular automata with interaction range $\ell\ge
3$. This is a family of integrable models for which there was no general theory so far.
We develop an algebraic framework for such models, generalizing known
methods from nearest neighbor interacting chains. This leads to a new integrability condition for medium range
Hamiltonians, which 
can be used to classify such models. A partial classification is performed in specific cases, including $U(1)$-symmetric
three site interacting models, and Hamiltonians that are relevant for interaction-round-a-face models. We find a number
of models which appear to be new. As an application we consider quantum brickwork circuits of various types, including
those that can accommodate the classical elementary cellular automata on light cone lattices.
In this family we find that the so-called Rule150 and Rule105 models are Yang-Baxter integrable with three
site interactions. We present integrable quantum
deformations of these models,
and derive a set of 
local conserved charges for them.
For the famous Rule54 model we find that it does not belong to the family of integrable three site models,
but we can not exclude Yang-Baxter integrability with longer interaction ranges.
\end{abstract}

\maketitle

\section{Introduction}

One dimensional quantum integrable models are special many body systems, which allow for exact solutions of their
dynamics. Their study goes back to the solutions of the Heisenberg spin chain by H. Bethe in 1931 \cite{Bethe-XXX} and
the exact 
treatment of the 2D classical Ising model by L. Onsager in 1944 \cite{Onsager-Ising}.  It was understood in the 60's
and 70's that a key element appearing in various types of quantum integrable models is the Yang-Baxter
equation, independently discovered by C. N. Yang  \cite{Yang-nested} and R. Baxter \cite{Baxter-Book}. 
A common algebraic framework was afterwards developed by the L. Faddeev and the Leningrad group (see for example the
historical review \cite{faddeev-history}). These elements of integrability connect seemingly different types of models,
such as the 2D integrable statistical physical models (for example the six vertex and eight vertex models), the famous
integrable spin chains such as the Heisenberg model or the Hubbard model, non-relativistic quantum gas models
\cite{lieb-liniger}, and integrable Quantum Field Theories (iQFT) \cite{mussardo-review}.

A re-occurring question over the decades has been the classification of (quantum) integrable models, together with an
attempt to give a precise definition of what integrability actually is. It appears that there is no single definition
encompassing all possible integrable models, but there are two very common elements appearing in integrable
models: the existence of a large set of extra conservation laws, and a completely elastic and factorized scattering of
the physical excitations \cite{caux-integrability}. Then the attempts for the classification can proceed along the lines
of either the charges  \cite{caux-integrability}, or by finding all possible factorized $S$-matrices as 
in iQFT \cite{mussardo-review}.

Focusing on quantum spin chains two big families of models have been studied extensively: those with nearest neighbor (n.n.)
interactions, and some long range models. In the latter case spins at an arbitrary distance can interact, although with strongly
decreasing coupling constants. The n.n. chains can be treated with the algebraic methods developed by the Leningrad
group \cite{faddeev-how-aba-works}, which are routinely used today. On the other hand, the treatment of the long range
chains is typically more involved. Examples are the Haldane-Shastry chain \cite{haldane-sashtry-1,haldane-sashtry-2} or
the Inozemtsev chain \cite{inozemtsev-chain} or long range version of the Hubbard model \cite{Hubbard-book}. In such
cases the construction of the exact eigenstates and also the charges is typically more complicated than in the
n.n. chains, see for example \cite{haldane-shastry-conserved,YB-longrange,Inozemtsev:2002vb}. 

Returning to the nearest neighbor chains, partial classifications have been achieved in certain specific
cases. Integrable nearest neighbor Hamiltonians are intimately connected with so-called regular solutions of the 
Yang-Baxter (YB) equation. Thus a classification can proceed by finding all solutions of the YB equation; this is
possible within restricted parameter spaces.
For example it was understood very early that solutions 
can be found by assuming underlying group or quantum group symmetries
\cite{kulish-sklyanin-YB,kulish-resh-sklyanin--fusion,jimbo-quantum-groups,Kennedy-all-su2-chains,Mutter-Schmidt-classification}.
However, the (quantum) group symmetric cases do not exhaust all possibilities,
and it is also desirable to find the remaining models. It
is not possible to list here all known integrable n.n. chains, therefore we just mention a few works that performed
classifications with certain restrictions \cite{classi1,classi2,classi3,classi4,classi5,Vieira-classification}.
More recently a systematic method was worked out in
\cite{marius-classification-1,marius-classification-2,marius-classification-3,marius-classification-4,sajat-lindblad},
based on well established ideas \cite{integrability-test} but leading to more detailed classifications than in prior
works. 

Despite all of this progress there is a class of models which has so far received relatively little attention:
Translationally invariant spin
chains where the Hamiltonian has a finite interaction range $\ell\ge 3$. We call these models
``medium range spin chains'' to distinguish them both from the nearest neighbor and the long range cases. It is
important that we are interested in models where the finite range Hamiltonian is the smallest dynamical charge, so we
dismiss those cases when the Hamiltonian is chosen as a linear combination of some of the short range charges of a
n.n. interacting chain. These cases can be interesting  on their own right (see for example \cite{Kluemper-GGE}), but
we are looking for models with genuinely new interactions. We also exclude those models, where the Hilbert space is a
constrained subspace of the usual tensor product space of the spin chains; an important example is the constrained XXZ
model and its generalizations treated for example in
\cite{constrained1,constrained5,constrained-multi,constrained2,constrained3,pronko-abarenkova-ising-limit,xxz-triple-point}
or the supersymmetric spin chains studied in \cite{susy-constrained-lattice-hopping1,susy-constrained-lattice-hopping2}. Such 
constraints are non-local, and we are looking for strictly local models with the usual tensor product Hilbert space. 

There are various reasons why the medium range models can be interesting. First of all there is the obvious academic interest: If
one was to uncover all possible integrable spin chains, then clearly one should consider these models as
well. We can expect that as we increase the interaction range, more and more possibilities open up, and a full
classification becomes less and less feasible in practice. Nevertheless it is desirable to develop a general theory
for such models, and at least some key ideas for the classification, which can be applied in restricted parameter
spaces. As a second motivation we can mention various recent research directions where medium range models were
encountered. 

\subsection{Medium range spin chains in the literature}

An early example for an interacting medium range spin chain is the Bariev-model, which has a three site Hamiltonian
\cite{bariev-model}. Generally the three site models can be pictured as a zig-zag spin
ladders, and this was also used in the presentation of the Bariev model. In this work we stick to the translationally invariant
representations. The algebraic explanation of the integrability of the Bariev model was given in
\cite{bariev-lax-1,bariev-lax-2}.

Recently there was interest in medium range chains which can be solved by free fermions of parafermions. A specific
three-site model was found in \cite{fendley-fermions-in-disguise} with generalizations studied later in
\cite{alcaraz-medium-fermion-1,alcaraz-medium-fermion-2,alcaraz-medium-fermion-3}. In these models the fermions are ``in
disguise'', which means they are not obtained by the usual Jordan-Wigner transformation. A more general theory for such
models was initiated in \cite{fermions-behind-the-disguise}.

A specific medium range model called the ``folded XXZ model'' was investigated in the recent works \cite{folded1,folded2,sajat-folded,folded-jammed}. There are two 
formulations of the model, and the dynamical Hamiltonian is a four-site or three-site operator, depending on the
formulation used. In the three-site formulation the model is seen as a special point of the Bariev model.  It was shown
in \cite{sajat-folded} that the model exhibits Hilbert space fragmentation, and it can be considered as a hard rod
deformation of the XX model.  Furthermore, its real time dynamics can be solved
exactly in certain special quench problems, thus the model can be considered as one of the simplest interacting spin
chains. It is remarkable that a spin chain with four site interactions has a simpler solution than the famous
XXZ chains. This shows that it is worthwhile to explore the medium range chains. 

In the recent work \cite{clifford1} a family of unitary transformations was studied which can generate a medium range
chain starting 
from a nearest neighbor model. The techniques applied here originate in quantum information theory,
and the transformations are members of the discrete Clifford group. 

\subsection{Integrable quantum circuits}

We should also mention the recent interest in integrable quantum circuits and classical cellular automata. This is a topic
closely connected to integrable Hamiltonians, and some of the models in the literature have medium range interactions.

The interest in quantum gate models is motivated in part by experimental advances, but also
by surprising theoretical results. For example it was found that simple (non-integrable) models with random unitary
gates can lead
to exact solutions, see for example
\cite{2015PhRvB..92b4301C,random-circuit-1,random-circuit-2,random-circuit-3,random-circuit-4,random-circuit-5,random-circuit-6}. In the
integrable setting there is interest for quantum circuits with unitary gates that span two, three of four sites (in the
following we will often use the short term ``unitary'' instead of ``unitary gate'').

First of all, a quantum circuit model with two-site
unitaries was developed in \cite{integrable-trotterization}: the model serves as an integrable Trotterization (discrete
time analog) of the XXZ spin chain. This idea goes back to the light cone regularization of integrable
QFT's \cite{DdV0,Volkov,Faddeev-Volkov}. More recently the same idea was also applied to dissipative systems
\cite{lindblad-circuit}. The key observation in these works is that the so-called $R$-matrix itself can be used as a
two-site quantum gate, and in certain cases this leads to discrete unitary time evolution with well defined
integrability properties. The original integrable spin chains can be recovered in the continuous time limit of the
quantum gate models.

An important family of quantum circuits  falls outside the realm of nearest neighbor
models. These are the elementary cellular automata on light cone lattices classified in \cite{rule54}, which can be considered
quantum circuits with special three site unitaries. The most important example is
 the Rule54 model, which is often called the simplest interacting integrable model \cite{rule54-review}. It is a
very special model with soliton-like behaviour, which allows for exact solutions
\cite{katja-bruno-lorenzo-rule54,katja-bruno-rule54-ghd,rule54-entangl} and integrable quantum deformations
\cite{vasseur-rule54} (see also \cite{sarang-rule54}). However, the connection with the standard Yang-Baxter
integrability remained unknown. 

Very recently a new algebraic framework was proposed for these cellular automata \cite{prosen-cellaut}, by making connection to the so-called Interaction Round-a-Face (IRF) models of statistical
physics (see for example \cite{RSOS-1,RSOS-2,IRFetiology,RSOS-H}). In  \cite{prosen-cellaut} new transfer matrices were developed for the classical cellular automata and
certain quantum 
deformations of them, and it was conjectured that the new construction gives new quasi-local conserved charges in these
models, thus proving their integrability. These IRF models are such that the three site quantum gates
have two control bits on the two sides and one action bit in the middle. If these IRF models are indeed integrable,
then they could serve as integrable Trotterization of some
spin chains with three site Hamiltonians, likely having a very similar structure. However, such connections have not yet
been found.

In the recent work \cite{sajat-cellaut} a cellular automaton was found with a four site update
rule, such that it can be considered as an integrable
Trotterization of the folded XXZ model mentioned above. However, the construction in \cite{sajat-cellaut} had some
drawbacks: it did not have space reflection symmetry, and the integrability was only proven for a certain
diagonal-to-diagonal transfer matrix, which is not adequate to treat the Cauchy problem. Nevertheless
the results of \cite{sajat-cellaut} indicate  that the cellular automata and medium range spin chains are indeed closely related,
and can be treated with the usual algebraic methods of integrability.

We should also note that there exists a family of classical cellular automata, where the integrability properties are
well understood: these are the so-called box ball systems \cite{box-ball,box-ball-review}. Here the update rules are
non-local in the sense that they can not be formulated using a simultaneous action of local update rules;
instead, they belong to the class of the so-called filter automata \cite{filter1,filter2,filter3,filter4}.
The box-ball models are noteworthy, because they display solitonic behaviour, and they are connected to a number of 
question in representation theory of quantum groups, combinatorics, and classical integrability
\cite{box-ball-review}. Recently the hydrodynamic behaviour of these models was also studied in \cite{box-ball-ghd}. In
our paper we do not treat these models, because we are interested in strictly local systems. 

\subsection{The goals of this paper}

Motivated by the findings discussed above, in this paper we strive towards a general theory for medium range integrable
models. It is our goal to 
develop the common algebraic structures, which will be generalizations of the known methods applied for nearest
neighbor models. We stress that up to now there has been no general framework for the medium range models, and even in those
cases where the algebraic background was developed (see for example the Bariev model \cite{bariev-lax-1,bariev-lax-2}),
it was based on ad hoc ideas lacking a general understanding. In contrast, in this work we establish the key relations
for the medium range models, focusing in particular on the three site and four site interacting cases. A special
emphasis will be put 
on the IRF models treated in \cite{prosen-cellaut}: we show that they can be embedded into our framework,
and we disprove some of the conjectures made in \cite{prosen-cellaut}. In particular, we argue that the Rule54 model is
not in the family of the three site interacting Yang-Baxter integrable models, but we do not exclude integrability with
longer interaction ranges.

In Section \ref{sec:main} we set the stage for our computations: we introduce the main concepts and also explain and
summarize some of our key results. The algebraic structures of integrability are then introduced in Section
\ref{sec:int}, which also includes our main results about the medium range spin chains. Integrable quantum circuits of
various types are constructed in Section \ref{sec:circuit}. The special class of models related to the elementary
cellular automata are considered in Section \ref{sec:IRFsec}. Here we also treat the results of
\cite{prosen-cellaut}. In Section \ref{sec:four} we study the four site interacting models.
Open questions are discussed in \ref{sec:disc}, and we present some of the technical computations
in the Appendices.

\section{Preliminaries}

\label{sec:main}

In this Section we introduce the key concepts regarding integrable spin chains, and we summarize our new results
for the medium range cases. We also introduce the quantum circuits, which can be considered as the discrete time
versions of the spin chain models. In this Section we avoid the algebraic treatment of the integrability properties,
instead we focus on the overall physical properties of these models, most importantly on the set of conserved charges.

First we introduce some notations that we use throughout the work, and afterwards we discuss the spin chains and the
quantum circuit models. The algebraic structures behind the integrability are presented later in Sections \ref{sec:int}
and \ref{sec:circuit}.

\subsection{Notations}

We consider homogeneous spin chains with translationally invariant Hamiltonians. The local Hilbert spaces are
$V_j=\complex^d$ with some fixed $d\ge 2$.  The full Hilbert space of the model in finite volume $L$ is
$\mathcal{H}=\otimes_{j=1}^L V_j$. The majority of our abstract results will not depend on the actual value of the local
dimension $d$, but in the concrete examples we consider $d=2$.

We say that an operator $\ordo(j)$ is local if its support is restricted to a limited
number of sites starting from $j$, such that the support does not grow with $L$ as we consider longer and longer
chains. We use notation $|\ordo(j)|$ for the range of the local operator (which we also call length). This means that
the support of $\ordo(j)$ with $|\ordo(j)|=\ell$ is the segment $[j,\dots,j+\ell-1]$. 

If a local operator has a fixed small range $\ell$, we will also use an alternative notation where we spell out the
sites on which it acts. For example for two-site operators we also write $\ordo_{j,j+1}$, for three site operators
$\ordo_{j,j+1,j+2}$, and so on. We will switch between the two notations depending on which is more convenient for the
actual computation. 

In our concrete examples we will treat spin-1/2 chains. In these cases we use the standard basis of the up and down
spins, but we will use the notations $\ket{\circ}=\ket{0}$ (empty site) and $\ket{\bullet}=\ket{1}$ (occupied site) for
these basis states, respectively. We use the 
standard Pauli matrices $\sigma^{x,y,z}$ and also the standard ladder operators $\sigma^\pm$, together with the
following projectors
onto the basis states:
\begin{equation}
  P^\circ=P^0=\frac{1+\sigma^z}{2},\qquad   P^\bullet=P^1=\frac{1-\sigma^z}{2}.
\end{equation}
An important operator that we will use often is the cyclic shift operator $\UU$ that translates the finite chain of
length $L$ to the right by one site. 

\subsection{Integrable spin chains with nearest neighbor interactions}

In the literature there is no single definition for quantum integrability. However, it is generally accepted that one of the
key properties of integrable models is the existence of a large set of additional conserved charges, which commute with
each other. Then the integrable
models can be classified according to the patterns of how these charges appear in the model \cite{caux-integrability}.

In this paper we focus on local spin chains, where the Hamiltonian is given by a strictly local Hamiltonian density:
\begin{equation}
  H=\sum_j h(j).
\end{equation}
Here $h(j)$ is a local operator with some range $\ell$. Periodic boundary conditions are understood throughout this
work.

The additional conserved charges are a set of operators $Q_\alpha$, where $\alpha$ is a label which we discuss below.
We require that each charge should be extensive with a local
density:
\begin{equation}
  Q_\alpha=\sum_j q_\alpha(j).
\end{equation}
In this work we restrict ourselves to these strictly local charges, even though it is known that in certain cases
so-called quasi-local charges also play an important role \cite{JS-CGGE,prosen-enej-quasi-local-review}. Furthermore,
there are integrable models where the 
charges typically grow super-extensively with the volume, see for example the discussion of the Haldane-Shastry chain in
\cite{caux-integrability}. Nevertheless, in this work  we focus only on the
extensive cases. 

It is generally required that the conserved charges commute
\begin{equation}
  [Q_\alpha,Q_\beta]=0,
\end{equation}
and the Hamiltonian should be a member of the set. The commutativity has to hold in every volume $L$ large enough so
that both charges fit into the system.

If these requirements are met, then generally the label $\alpha$ can be chosen simply as the length of the charge
density, therefore we will use the convention throughout this work
\begin{equation}
  |q_\alpha(j)|=\alpha.
\end{equation}

Within this class of models the most studied ones are nearest neighbor interacting ones, thus
we can identify $H=Q_2$. In such cases the allowed values for $\alpha$ are the integers starting from 2. If there is
also a global $U(1)$-symmetry then $\alpha=1$ is also allowed.

Within this class the most important cases are the spin-1/2 chains, for which a full classification 
(including non-Hermitian cases) was performed in
\cite{marius-classification-1,marius-classification-2}. We do not treat the classification here, instead we just mention
the most important examples. 

Let us start with a generic (non-integrable) spin-1/2 chain and let us require space reflection invariance. Then it can
be shown that the most general nearest neighbor Hamiltonian (apart from global $SU(2)$ rotations) is the XYZ model with
magnetic fields:
\begin{multline}
  \label{XYZh}
  H=\sum_j \left( J_x \sigma^x_{j}\sigma^x_{j+1}+J_y \sigma^y_{j}\sigma^y_{j+1}+J_z \sigma^z_{j}\sigma^z_{j+1}\right.+\\
+\left.    h_x \sigma^x_j+h_y \sigma^y_j+h_z \sigma^z_j\right).
\end{multline}
This model is not integrable except for special cases \cite{xyz-not-int}.

One integrable family is when $h_x=h_y=h_z=0$, and the couplings $J_{x,y,z}$ are arbitrary, this is known as the XYZ
model. A special point with $U(1)$ symmetry is the XXZ chain with $J_x=J_y$, which allows for a non-zero $h_z$. A
further special point is the $SU(2)$ invariant Heisenberg spin chain with equal couplings, which allows for arbitrary
magnetic fields. 

An other special integrable family within \eqref{XYZh} are the so-called XYh models, where $J_z=0$ and
$h_x=h_y=0$. A  special model of this family is the quantum Ising chain where also $J_y=0$. The XYh family can be solved
by free 
fermion techniques \cite{XYh-Fan-Wu}.

\subsection{The Reshetikhin condition}

A common property of the nearest neighbor models is that they satisfy the so-called Reshetikhin condition of
integrability. There are 
multiple formulations of this condition, and now we review the most general one. The algebraic background is treated
later in Section \ref{sec:int}.

First of all we quote a Conjecture that was presented in \cite{integrability-test}. We put this in a slightly modified
form:

\begin{conjecture}
  \label{conj:1}
  A nearest neighbor spin chain with a
dynamical Hamiltonian $H$ is integrable iff there exists an extensive 3-site charge $Q_3$ which is functionally
independent from $H$ and the possible one site charges of the model, and which commutes with $H$ for every volume $L\ge
3$.
\end{conjecture}

As far as we know no counter-examples have been found so far, but the Conjecture has not yet been proven
either. It is important that we added the condition that the Hamiltonian should be dynamical: If this condition is not
satisfied, then simple counter examples can be found, see the discussion in Appendix \ref{sec:counter}.

Now we also present the Reshetikhin condition in a rather general form:

\begin{conjecture}
  \label{conj:2}
If there exists a conserved charge $Q_3$ commuting with the dynamical two-site Hamiltonian $H=\sum_j h(j)$, then it can
be written as $Q_3=\sum_j q_3(j)$ with
\begin{equation}
  \label{q3f}
  q_3(j)=[h(j),h(j+1)]+\tilde h(j),
\end{equation}
where $\tilde h(j)$ is also a two-site operator.  
\end{conjecture}

It might appear that the Conjecture allows for a lot of freedom due to the presence of  $\tilde h(j)$, but the fact that
this operator has to be a two-site operator is rather restrictive.

The conjecture can be proven backwards: If the spin chain is Yang-Baxter integrable, then the precise form of the
density $q_3(j)$ follows from the underlying algebraic objects, and it has precisely the form \eqref{q3f}. However, it
is not known how to prove the Conjecture without assuming Yang-Baxter integrability. We put forward that  $\tilde
h(j)=0$  in models where the so-called $R$-matrix is of {\it difference form}; this corresponds to the original
Reshetikhin condition treated in \cite{integrability-test}. On the other hand, 
 $\tilde h(j)\ne 0$  in other models such as the Hubbard model \cite{hubbard-boost}. 

Conjecture \ref{conj:2} was used in the recent works
\cite{marius-classification-1,marius-classification-2,marius-classification-3,marius-classification-4,sajat-lindblad}
for the classification of integrable nearest neighbor chains in various circumstances. We will not review this
classification here, instead we will focus on the generalization of the Reshetikhin condition to medium range spin chains.

\subsection{Medium range spin chains}

In this work we treat integrable spin chains and closely related quantum gate models where the interaction range is
$\ell\ge 3$. We call these models ``medium range chains''. We propose the following definition:

\begin{definition}
   A medium range integrable spin chain is a model which has an infinite set of commuting local
charges $\{Q_\alpha\}$ where $\alpha\in \mathcal{S}$ with $\mathcal{S}\subset \egesz^+$, such that the lowest dynamical
charge (which is regarded as the Hamiltonian) has a range $\ell \ge 3$.  
\end{definition}

Once again it is important to add the requirement of having a dynamical charge as the Hamiltonian. A good example for
the usefulness of this requirement is the so-called ``folded XXZ model'' treated in \cite{folded1,folded2,sajat-folded},
where the first four charges are
\begin{equation}
  \label{foldedQ}
   \begin{split}
       Q_1&=\sum_{j=1}^L \sigma^z_j,\qquad
    Q_2=\sum_{j=1}^L \sigma^z_j\sigma^z_{j+1},\\
   Q_3&=\sum_j
i (\sigma^z_j+\sigma^z_{j+3})(\sigma^+_{j+1}\sigma^-_{j+2}-\sigma^-_{j+1}\sigma^+_{j+2}),\\
  Q_4&=\sum_{j=1}^L (1+\sigma^z_j\sigma^z_{j+3})( \sigma^+_{j+1}\sigma^-_{j+2}+ \sigma^-_{j+1}\sigma^+_{j+2}).
  \end{split}
\end{equation}
Based on the existence of the charge $Q_2$ one could regard this model as nearest neighbor interacting, but $Q_2$ does
not generate dynamics. Instead, in this model $Q_4$ is regarded as the Hamiltonian, because $Q_4$ is the first parity
symmetric dynamical charge of the model.

\subsection{Three site models -- general remarks}

\begin{figure}
\centering
\includegraphics[width=0.7\columnwidth]{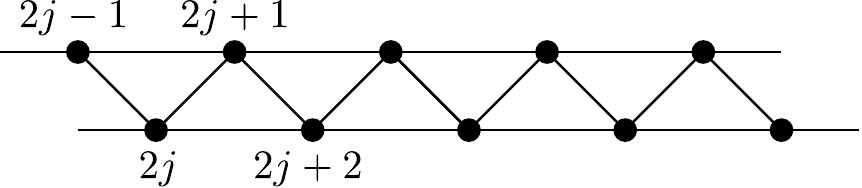}
\caption{Graphical illustration of a zigzag spin ladder. It is natural to expect three site interactions for each
  plaquet.}
\label{fig:zigzag}
\end{figure}

Let us focus on the case with $\ell=3$. In this case we identify $H=Q_3$ and we also write
\begin{equation}
  \label{HQ3}
  H=\sum_j  h_{j,j+1,j+2},
\end{equation}
where $h_{j,j+1,j+2}$ is the three site Hamiltonian density.

Such models can be interpreted very naturally as a zig-zag spin ladder. See figure \ref{fig:zigzag}. However, we will focus on the
translationally invariant representation \eqref{HQ3}.

Our main results are finding the general algebraic structures behind the three site models, which lead to a
generalization of the Reshetikhin condition. We formulate the following conjecture:

\begin{conjecture}
  \label{conj:q5}
A three site Hamiltonian is integrable, iff the charge $Q_5=\sum_j q_5(j)$ defined by
\begin{multline}
  \label{q5d}
  q_5(j)=[h_{j,j+1,j+2},h_{j+1,j+2,j+3}+h_{j+2,j+3,j+4}]+\\ 
+ \tilde h_{j,j+1,j+2}
\end{multline}
commutes with the Hamiltonian in every volume $L\ge 5$.  Here $\tilde h_{j,j+1,j+2}$ is an other three-site operator.
\end{conjecture}
Clearly, this is a generalization of Conjecture \ref{conj:2}. The construction of $Q_5$ through \eqref{q5d} is one of our key
results. Its derivation from an underlying algebraic theory is presented in Section \ref{sec:Q5}.  We stress again that
this result is very restrictive: it uses two three-site operators $h_{j,j+1,j+2}$  and $\tilde h_{j,j+1,j+2}$  to
generate a five site charge. 

Based on this result it is possible to perform a classification of integrable spin chains with 3-site interactions. This
is rather analogous to the ideas used in \cite{Kennedy-all-su2-chains,Mutter-Schmidt-classification} and later in
\cite{marius-classification-1,marius-classification-2,marius-classification-3,marius-classification-4,sajat-lindblad}.
The
idea is to make an Ansatz for $h_{j,j+1,j+2}$ and $\tilde h_{j,j+1,j+2}$, possibly including a number of free
parameters, to construct $Q_5$ using 
\eqref{q5d} and to check the commutation relation $[H,Q_5]=0$ on spin chains with medium length. We implemented this
strategy using the program \texttt{Mathematica} and performed partial classifications on spin 1/2 chains. The generic density
$h_{j,j+1,j+2}$ has a total number of $8^2=64$ parameters, which is a too large parameter space for practical
computations. Therefore we performed partial classifications along restricted subspaces which can bear physical
relevance. The results are presented in Subsections \ref{sec:classification} and \ref{sec:IRFH}.

\subsection{Quantum gates and cellular automata}

\label{sec:qgatesintro}

We also consider brickwork type quantum circuits, where the fundamental local unitaries have a support of
$\ell$ sites. The interest in such models is manifold: On the one hand, they can be understood as integrable models
in discrete time, or alternatively as integrable Trotterizations of the continuous time spin chain
models. On the other hand, they are interesting because they can lead to classical cellular automata, thus presenting
one more link between the worlds of quantum and classical integrability. Finally, they are also relevant to quantum
computing and real world experiments.

Let us sketch the general schemes behind our quantum circuit models. First we present the abstract formulation of the
update rules, and we give more concrete examples later. It is necessary to start with the most general form, because
later in this work we consider multiple types of constructions.

We build systems with Floquet-type discrete time
evolution, such that the equal time update rules are given by local unitaries. Let us fix an interaction range $\ell$,
and consider a local unitary $U^{(\ell)}(j)$ which acts on the segment $[j,\dots,j+\ell-1]$ of the spin chain. Typically
these unitaries will also depend on a continuous parameter $u$ (the spectral parameter) and on a small number of extra
parameters characterizing a family of models. Then we construct a brickwork type update rule for the whole spin chain
using the single unitaries. We build a Floquet-type cycle with time period $\tau$:
\begin{equation}
  \label{floquet}
  \mathcal{V}=\mathcal{V}_\tau\dots\mathcal{V}_1,
\end{equation}
such that each update step $\mathcal{V}_j$ with $j=1\dots \tau$ is a product of commuting local unitaries acting at the
same time. We formalize it as
\begin{equation}
  \label{Vj}
  \mathcal{V}_l=\prod_{k}  U^{(\ell)}(x_k+\Delta_l).
\end{equation}
Here $x_k$ are coordinates that specify the placement  of the local unitaries within a single time update, and
$\Delta_l$ is a displacement which depends on the discrete time index $l$, signaling the position within the
Floquet-type cycle. 

The local unitaries within a given time step should commute with each other, such that the order of the product
\eqref{Vj} does not matter. This requirement can be important for practical purposes (implementation of the quantum
circuits in experiments), but also for theoretical reasons. In the most general case the commutativity holds if
the supports are non-overlapping, i.e. $x_{k+1}\ge x_k+\ell$. Nevertheless the supports can have a non-zero overlap, if the
commutativity is guaranteed by other means, for example if the local unitaries act diagonally on the overlapping
sites.

The simplest example for the brickwork construction discussed above is  the alternating circuit discussed for example in
\cite{integrable-trotterization}. In this case $\tau=2$, we can choose $x_k=2k$, and $\Delta_l=l$ with $l=1,2$. For a
graphical representation see Figure \ref{fig:2siteQgates} later in Section \ref{sec:circuit}. We can build similar
structures with interaction range 
$\ell=3$, the most obvious choice is $\tau=3$, $x_k=3k$, and $\Delta_l=l$ with $l=1,2,3$ (see Figure
\ref{fig:3siteDiag2Diag}). Such a brickwork circuit was introduced in 
\cite{sajat-cellaut}.
Later we will also present other types of constructions. 
 
As mentioned earlier, the local unitaries typically have a spectral parameter $u$ which can be tuned freely. This means
that we can build families of quantum circuits. These families can have special points with very specific physical behaviour.

For example, in the most typical case the unitaries become equal to the identity at
the special point $u=0$. Furthermore, the first order expansion in $u$ gives a local Hamiltonian
$h$ with interaction range $\ell$. Assuming $u\in\valos$ this is formalized as
\begin{equation}
  U^{(\ell)}(u|j)=1+iu h_{j,\dots,j+\ell-1}+\ordo(u^2).
\end{equation}
In such a case the Floquet circle $\mathcal{V}$ can act as a Trotterization of the global Hamiltonian $H$:
\begin{equation}
  \mathcal{V}=1+iuc H+\ordo(u^2),
\end{equation}
where $c$ is a real number that depends on the details of the construction, such as the Floquet period $\tau$, the coordinate
differences $x_{k+1}-x_k$ and the displacements $\Delta_l$ in \eqref{Vj}.

We intend to construct quantum circuits with precisely such a behaviour, so that $H$ is one of our medium range
integrable Hamiltonians. Furthermore we require that the quantum circuit itself should have certain integrability
properties. These depend on the particular construction, and will be discussed in detail in
\ref{sec:circuit}; the local unitaries will be derived from the so-called Lax operators
of the medium range model. 

In some models the local unitaries become deterministic for a special value $u$ of the rapidity parameter. This means
that $U^{(\ell)}(j)$ simply just permutes the states in the computational basis (possibly with some phases added),
without creating any linear combinations. The states of the computational basis can be considered {\it classical}, because
every spin has a fixed value; if the local unitaries are deterministic then classical states are mapped to classical
states during time evolution. This means that the quantum circuit can be considered a {\it classical cellular automaton} at
this special point $u$. An integrable example for such model was presented recently in \cite{sajat-cellaut}. An
other important class for such models are the Interaction-Round-A-Face models treated in \cite{prosen-cellaut},
which lead to the elementary cellular automata on light-cone lattices. In the Subsection below we discuss these models
in detail. 

\subsection{Elementary cellular automata on light cone lattices}

\label{sec:cellintro}

\begin{figure}
\centering
\includegraphics[width=0.85\columnwidth]{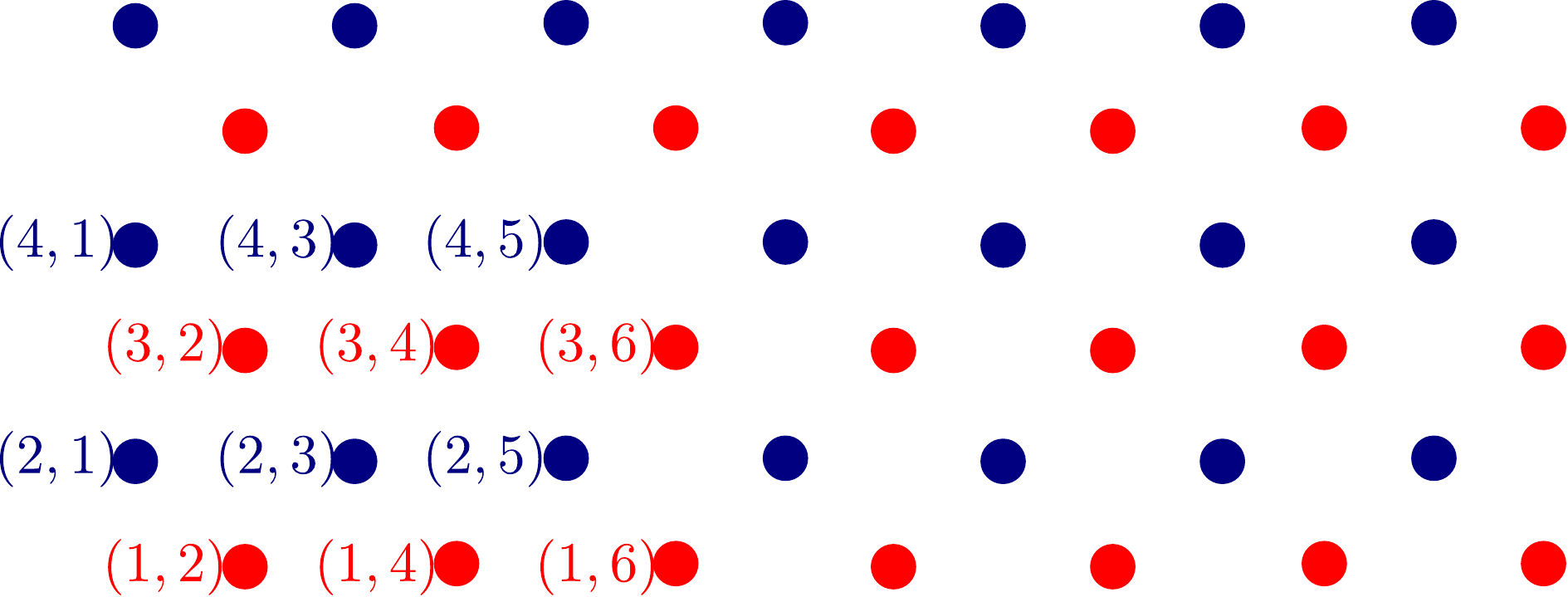}
\caption{The light cone lattice for the elementary cellular automata. We assign coordinates $(t,x)$ to the sites, where
  $t$ is interpreted as a time variable. The $x$ coordinates are increased by steps of 2, and they take even (odd)
  values if the $t$ coordinate is odd (even), respectively. The update of the cellular automata proceeds in the time
  direction upwards, and each spin is given a new value using the state of 3 of its neighbors (to the directions
  South-West, South, and South-East). For example, the spin at position $(3,2)$ is given a value using the spins at
  $(2,1)$, $(1,2)$ and $(2,3)$.
  }
\label{fig:rombus}
\end{figure}

These are classical 2-state models where the variables are defined on a light cone lattice, see
Fig. \ref{fig:rombus}. The vertical 
dimension is interpreted as the direction of time. Each cell is updated depending on the state of
its 3 neighbors (to the South-west, South, and south-East), and the update rules are homogeneous both in space and
time.
Accordingly, there are $2^{2^3}=256$ such models, and they have been studied and classified in the seminal works
\cite{cellaut-class,rule54}. The nomenclature for these models follows the original proposal of Wolfram
\cite{wolfram-cellaut1} (see also \cite{rule54}).
Recently these cellular automata  attracted considerable attention due to the integrability properties
of some of its members: the Rule54, Rule150 and Rule201 models (see the review \cite{rule54-review} and
also \cite{rule150-markov,rule201} ).

As the update rules depend on 3 bits, it is tempting to look for a connection with our 3-site interacting Hamiltonians
and quantum gate models. Therefore we provide here a formulation of the models which fits into the
framework given above. Afterwards we provide a simple classification of the physically interesting
models. Finally we summarize our main results, with the technical computation presented later in Section \ref{sec:circuit}.

\begin{figure}
\centering
\includegraphics[width=1.0\columnwidth]{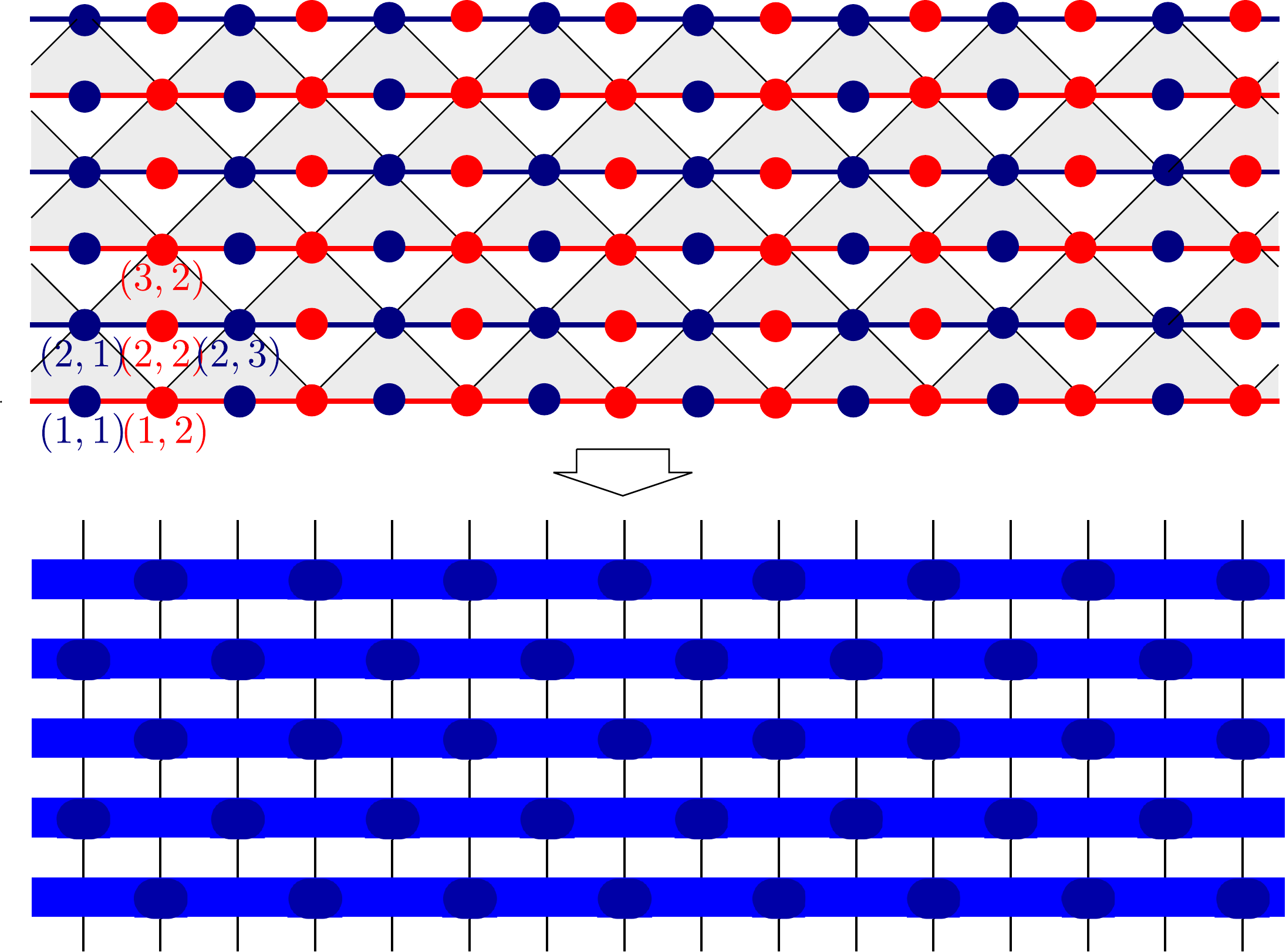}
\caption{Two different representations for the cellular automata on the light cone lattices. -- On the top we
 complement the original arrangement into a rectangular lattice, by adding new sites to the centers of the
  faces of the original lattice. The spins at the newly added sites are taken to be identical with those immediately
  below them, for example the state of the new site at $(2,2)$ is copied from the spin at $(1,2)$. This way we can
  formulate an update rule using three site unitaries on a rectangular lattice. For example the spin at $(3,2)$ is given
a value using the three spins $(2,1)$, $(2,2)$ and $(2,3)$ from the previous row. The gray triangles show the direction
of the update steps. -- On the bottom we show a representation using quantum gates. Now the spin variables live on the
vertices and three site unitaries update them. The unitary gates are placed so that the neighboring pairs overlap at a
control bit, which are shown by the darker shaded circles.}
\label{fig:rect}
\end{figure}

First we transform the light cone lattice into a regular rectangular lattice. The idea is to add new sites to the
light cone lattice to the centers of the faces, see Fig. \ref{fig:rect}.
Then on this rectangular lattice we formulate a Floquet-type update rule
with period $\tau=2$, such that at each step the odd or even sites are updated, respectively. In this formulation the
local update is performed by a three-site $U^{(3)}$ which is actually deterministic in the computational
basis. Furthermore, it has a structure which was discussed already for the IRF type Hamiltonians above: The quantum gate
acts diagonally on the first and last bits, whereas it has an action bit in the middle. In the particular case we have
\begin{equation}
  \label{IRFU}
  U^{(3)}(j)=\sum_{a,b=0,1} P_j^a f^{ab}_{j+1}  P^b_{j+2},
\end{equation}
where $P^a_j$ is a projector to basis state $a$ acting on site $j$, and $f^{ab}_{j+1}$ is a collection of 4 matrices
acting on site $j+1$. Then the full update rule is specified by
\begin{equation}
  \label{IRFV}
  \mathcal{V}=\mathcal{V}_2\mathcal{V}_1
\end{equation}
with
\begin{equation}
  \label{IRFVj}
  \mathcal{V}_l=\prod_{k=1}^{L/2} U^{(3)}(2k+l),\qquad l=1,2.
\end{equation}
The neighboring three site unitaries overlap at one site, but they commute because they act diagonally on the boundary
sites. Thus \eqref{IRFVj} is well defined without specifying the order of the action the quantum gates.

Let us now discuss the classification of these models. We wish to have deterministic time evolution (without phases),
and we also require time 
reversibility, which follows from the requirement of unitary. Therefore
the only possibilities are 
\begin{equation}
  \label{fabch}
  f^{ab}=1\quad\text{or}\quad f^{ab}=\sigma^x.
\end{equation}

The choices \eqref{fabch} leave us
with $2^4=16$  models. If we also require space reflection invariance we end up with $2^3=8$ models.
Those two models where all $f^{ab}$ matrices are identical are completely
trivial, which leaves us with 6 models. Out of these 6 models we can choose the following 4, which are not related to
each other by overall spin reflection \cite{prosen-cellaut}:

\begin{equation}
\label{eq:rule54}
\text{Rule54:  } f^{00}=1, \quad f^{01}=f^{10}=f^{11}=\sigma^x  
\end{equation}

\begin{equation}
   \label{eq:rule105}
\text{Rule105: }  f^{00}=f^{11}=\sigma^x, \quad f^{01}=f^{10}=1
\end{equation}

\begin{equation}
   \label{eq:rule150}
\text{Rule150: }  f^{00}=f^{11}=1, \quad f^{01}=f^{10}=\sigma^x
\end{equation}

\begin{equation}
  \label{eq:rule201}
\text{Rule201:  }  f^{00}=\sigma^x,\quad  f^{01}=f^{10}=f^{11}=1.
\end{equation}

Additional two new models can be obtained by a complete spin reflection of the Rule54 and Rule201 models.

The Rule105 and Rule150 models are not completely independent either, because their $f$-matrices are obtained from each
other by a multiplication with $\sigma^x$. This means that
\begin{equation}
  \mathcal{V}_k^{(150)}=X_k  \mathcal{V}_k^{(105)}= \mathcal{V}_k^{(105)} X_k, \qquad k=1,2,
\end{equation}
where we defined
\begin{equation}
  X_k=\prod_{j=1}^{L/2} \sigma^x_{2j+k}, \qquad k=1,2.
\end{equation}
We can also observe the commutation relations
\begin{equation}
[  \mathcal{V}_1^{(105)},X_2]=[  \mathcal{V}_2^{(105)},X_1]=0,
\end{equation}
and similarly for the operators of the Rule150 model. Altogether this implies that the combined Floquet steps are
related as
\begin{equation}
   \mathcal{V}^{(150)}=X  \mathcal{V}^{(105)},
\end{equation}
where $X$ is the global spin reflection operator given by
\begin{equation}
  X=X_2X_1=\prod_{j=1}^{L} \sigma^x_{j}.
\end{equation}
From these relations we can also derive
\begin{equation}
  \left( \mathcal{V}^{(150)}\right)^2=\left( \mathcal{V}^{(105)}\right)^2.
\end{equation}
Thus the physical behaviour of the two models could be considered the same, up to a staggered global spin
reflection.

An important property of these models is that the two different possibilities for the Floquet time step
are actually inverses of each other:
\begin{equation}
  \label{Vinv}
  \mathcal{V}_1\mathcal{V}_2=\left(\mathcal{V}_2\mathcal{V}_1\right)^{-1}
\end{equation}
which follows simply from
\begin{equation}
  (\mathcal{V}_1)^2=(\mathcal{V}_2)^2=1.
\end{equation}
As a result the two Floquet operators actually commute
with each other:
\begin{equation}
  \label{Vcomm}
  [\mathcal{V}_1\mathcal{V}_2,\mathcal{V}_2\mathcal{V}_1]=
  [\mathcal{V}_1\mathcal{V}_2,\UU^{-1}\mathcal{V}_1\mathcal{V}_2\UU]
  =0.
\end{equation}

The four models listed above have been studied in a number of works recently (see the review \cite{rule54-review} for
the Rule54 
model and also \cite{rule150-markov,rule201} for the Rule150 and Rule201 models), nevertheless the algebraic background
for 
their observed integrability properties was not understood for a long time. Recently a new framework was introduced in
\cite{prosen-cellaut} which claimed to solve this problem. We also treat these cellular automata; our main results are as follows.

We find that the Rule105 and Rule150 models can be embedded into our framework of three-site interacting models.
We find a rapidity dependent family of commuting transfer matrices which includes the time evolution operator of the model as
a particular case. This family of transfer matrices is regular, and we derive a set of local conserved charges that
commute with the discrete time evolution operator of the cellular automata. Even though we use a different formalism, we show
explicitly that our construction is identical to that one of \cite{prosen-cellaut} for these special class of models.

On the other hand, for the Rule54 and Rule201 we find that they are not in the family of three site
interacting integrable models. In the case of the Rule54 model we also show that the transfer matrices of
\cite{prosen-cellaut} are 
algebraically dependent on three known conserved charges of the model, therefore those transfer matrices do not yield new
information; this is presented in Subsection \ref{sec:prosen}.

\section{Integrability structures}

\label{sec:int}

In this Section we present the integrability structures behind the spin chains under consideration. We start with a
brief treatment of the nearest neighbor chains, and afterwards we turn to our new results.

\subsection{Nearest neighbor interacting spin chains}

We construct a transfer matrix which serves as a generating function for the conserved charges of the model under
consideration. For an introduction into the methods we refer the reader to
\cite{faddeev-how-aba-works,Korepin-Book}. 

First we construct the monodromy matrix as follows. We take an auxiliary space $\complex^{d'}$.
 The Lax operator $\La_{a,j}(u)$ acts on the tensor product of the auxiliary space (denoted by the index $a$) and a physical space
with site index $j$.  The variable $u$ is called spectral parameter.  In those cases when the transfer matrix generates
the Hamiltonian and the other charges we have $d'=d$.

\begin{figure}
\centering
\includegraphics[width=0.6\columnwidth]{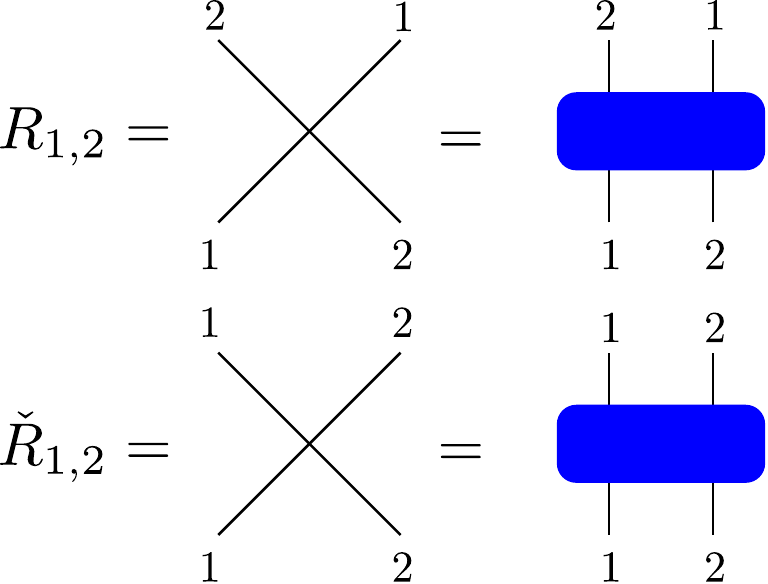}
\caption{Graphical illustration of the operators $\check R_{1,2}$ and $R_{1,2}$; we suppressed the dependence on the
  rapidity parameters. On the left we show the usual notation which comes from the vertex models. On the right we depict
the same operators as quantum gates. The two matrices differ in a permutation, and in this notation $\check R_{1,2}$
acts such that the two vector spaces are kept in place.}
\label{fig:Rconv}
\end{figure}

The monodromy matrix for a finite volume $L$ is defined as
\begin{equation}
  M_a(u)=   \La_{a,L}(u)\dots \La_{a,2}(u)\La_{a,1}(u),
\end{equation}
and the transfer matrix is
\begin{equation}
  \label{teredeti}
  t(u)=\text{Tr}_a\ M_a(u).
\end{equation}
The transfer matrices  form a commuting set of operators if the Lax operators satisfy the following
exchange relations, where $a$ and $b$ denote two auxiliary spaces:
\begin{multline}\label{RLL}
  R_{b,a}(\nu,\mu)  \mathcal{L}_{b,j}(\nu)  \mathcal{L}_{a,j}(\mu)=\\= \mathcal{L}_{a,j}(\mu) \mathcal{L}_{b,j}(\nu)    R_{b,a}(\nu,\mu).
\end{multline}
Here $R(u,v)$ is the so-called $R$-matrix which satisfies the Yang-Baxter relations
\begin{multline}
  \label{YB}
     R_{12}(\lambda_{1},\lambda_2)R_{13}(\lambda_1,\lambda_3)R_{23}(\lambda_2,\lambda_3)=\\
  =R_{23}(\lambda_2,\lambda_3) R_{13}(\lambda_1,\lambda_3) R_{12}(\lambda_{1},\lambda_2).  
\end{multline}
The figure \ref{fig:Rconv} shows the graphical presentation of the $R$-matrix.
In the models of physical relevance the $R$-matrix also satisfies the so-called regularity property
\begin{equation}
  \label{eq:regular}
  R_{ab}(u,u)\sim \Pe_{ab},
\end{equation}
where $\Pe_{ab}$ is the permutation operator acting on the tensor product space.

If the regularity condition holds then the following inversion can be established:
\begin{equation}
  \label{Rinv}
  R_{12}(\lambda,\mu)R_{21}(\mu,\lambda)\sim 1,
\end{equation}
where
\begin{equation}
 R_{21}(\mu,\lambda)=\Pe R_{12}(\mu,\lambda)\Pe.
\end{equation}
For the sake of completeness we present the derivation of \eqref{Rinv} using the Yang-Baxter equations and
\eqref{eq:regular} in Appendix \ref{sec:inv}. 

\begin{figure}
\centering
\includegraphics[width=0.8\columnwidth]{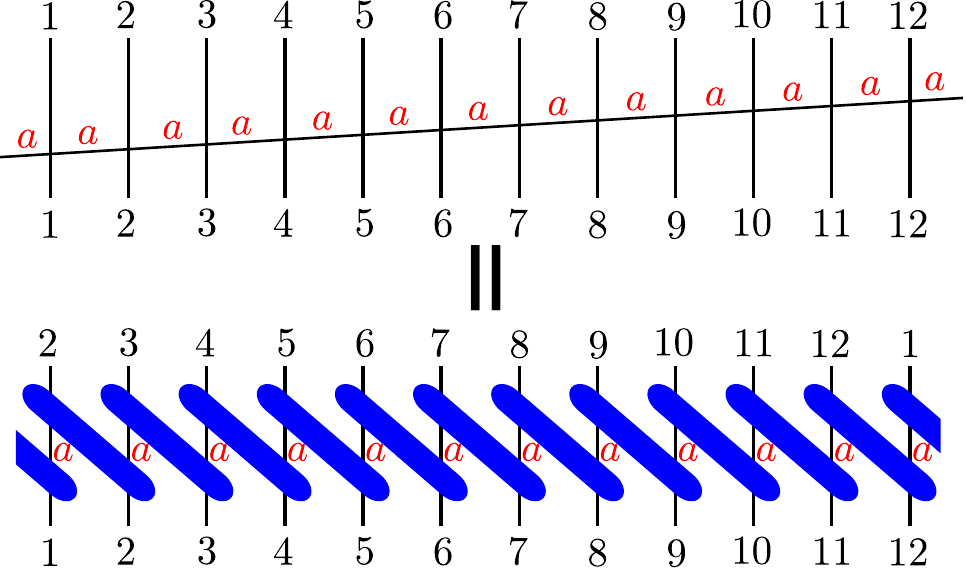}
\caption{Graphical illustration of transfer matrix \eqref{eq:transfer}. Here $a$ denotes the auxiliary space. The first
  representation is the usual from the literature, which shows the transfer matrix as a concatenation of Lax
  operators. The second picture shows the same object using a succession of quantum gates. The two representations are
  not very different, but generalizations of the second one will be more convenient in the more complicated cases that
  will follow.}
\label{fig:transfer}
\end{figure}

\begin{figure}
\centering
\includegraphics[width=0.8\columnwidth]{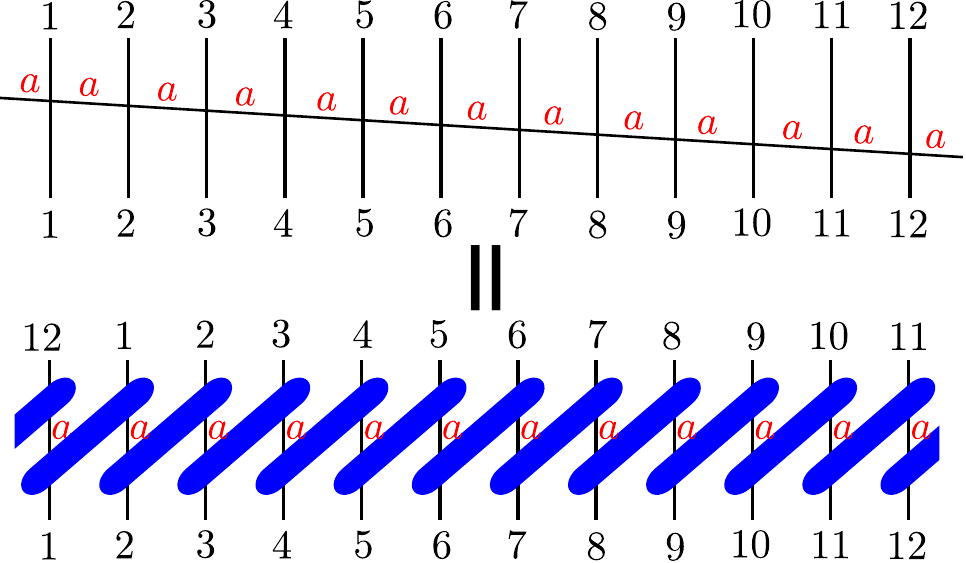}
\caption{Graphical illustration of transfer matrix \eqref{eq:transferRev}.}
\label{fig:transferRev}
\end{figure}

One can use the $R$-matrix itself as a Lax operator:
\begin{equation}
  \label{xi0}
\mathcal{L}_{a,j}(\mu)=R_{a,j}(\mu,\xi_0),
\end{equation}
where $\xi_0$ is a fixed parameter of the model. In such a case the YB relation is equivalent to the RLL relation and the transfer matrix reads as
\begin{equation} \label{eq:transfer}
 t(u) = \mathrm{Tr}_a R_{a,L}(u,\xi_0) \dots R_{a,2}(u,\xi_0) R_{a,1}(u,\xi_0).
\end{equation}
There is an alternative way to define a transfer matrix, where the ordering of the sites is the opposite, and the role
of the auxiliary and physical spaces is exchanged:
\begin{equation}
  \label{eq:transferRev}
  \bar t(u) = \mathrm{Tr}_a R_{1,a}(\xi_0,u) R_{2,a}(\xi_0,u)  \dots R_{L,a}(\xi_0,u).
\end{equation}
It can be proven using the Yang-Baxter relation that
\begin{equation}
  [t(u),\bar t(v)]=0.
\end{equation}
In some of our constructions below it will be more convenient to use \eqref{eq:transferRev} instead of
\eqref{eq:transfer}.
In figure \ref{fig:transfer} and \ref{fig:transferRev}  we can see the graphical presentations of these two transfer matrices.

The regularity condition ensures that at the special point $\xi_0$ we have the initial conditions
\begin{equation}
  t(\xi_0)=\UU, \qquad \bar{t}(\xi_0)=\UU^{-1}
\end{equation}
where $\UU$ is the cyclic shift operator on the chain.

A commuting set of local charges is then constructed as
\begin{equation}
  \label{Qalphadef}
  Q_\alpha\quad\sim\quad \left.(\partial_u)^{\alpha-1}\log(t(u)) \right|_{u=\xi_0}.
\end{equation}
In many models the $R$-matrix is of difference form, which means
\begin{equation}
  R(u,v)=R(u-v).
\end{equation}
In such a case the parameter $\xi_0$ is irrelevant, and it is conventional to set it to $\xi_0=0$. However, we will
consider the generic case where the $R$-matrix is not necessarily of difference form.

Let us investigate the first few charges in detail. Performing the first derivative of \eqref{Qalphadef} we get the
nearest neighbor Hamiltonian
\begin{equation}
  H=Q_2=\sum h_{j,j+1}
\end{equation}
with
\begin{equation}
  h_{j,j+1}=\left.\partial_u \check \La_{j,j+1}(u)\right|_{u=\xi_0},
\end{equation}
where
\begin{equation}
  \check \La_{j,j+1}(u)=\Pe_{j,j+1} R_{j,j+1}(u,\xi_0).
\end{equation}
For this operation it is crucial that the regularity condition is satisfied. 

For the third charge we get the general expression
\begin{equation}
  \label{Q3form1}
  Q_3=\sum_j [h_{j,j+1},h_{j+1,j+2}]+    \check \La''_{j,j+1}(\xi_0)- h^2_{j,j+1}.
\end{equation}
We can see that in addition to the commutator we have two-site operators, and altogether this expression takes the form
\eqref{q3f} announced earlier.

Assuming that the $R$-matrix has difference form and setting $\xi_0=0$ then the following inversion relation holds:
\begin{equation}
  \label{Linversion}
  \check \La(u)\check \La(-u)=1.
\end{equation}
This implies that the last two terms in \eqref{Q3form1} cancel, which can be seen after taking the second derivatives of
\eqref{Linversion} with respect to $u$ and substituting $u=0$. In this case we end up with
\begin{equation}
  \label{Q3form2}
  Q_3=\sum_j [h_{j,j+1},h_{j+1,j+2}].
\end{equation}
We can see that \eqref{Q3form2} follows from \eqref{Rinv} if the $R$-matrix is of difference form. 
However, in
other cases it does not hold generally. Perhaps the most famous example where it does not hold is the Hubbard model and
its inhomogeneous versions \cite{Murakami-Hubbard-inhom}.

The form \eqref{Q3form1} yields the original version of the Reshetikhin condition used for example in
\cite{integrability-test}. 

\subsection{Three site interactions}

\label{sec:threesite}

Let us now consider a three site interacting model, defined by the Hamiltonian
\begin{equation}
\label{H3}
  H=\sum_j h_{j,j+1,j+2}.
\end{equation}
Below we conjecture a generic form of the Lax operator for such models. We motivate our conjecture by some simple
ideas. We do not assume the existence of a proper translationally invariant Lax operator from the start, instead we
develop arguments that motivate its existence and its properties.

The key idea is to build a nearest neighbor chain out of our model by grouping together (or gluing) every two spins into
blocks.
Therefore we consider our spin chain in an even volume $L=2k$.
We pair the spins, and label the pairs using the original coordinates, for example $(j,j+1)$. Then we obtain a new model
with $k$ sites and local dimension 
$d^2$. The Hamiltonian for the new model can be written as 
\begin{equation}
  \tilde H=\sum_{j=1}^{L/2} \tilde h_{j,j+1},
\end{equation}
where we can choose for example
\begin{equation}
  \tilde h_{j,j+1}=h_{2j,2j+1,2j+2}+h_{2j+1,2j+2,2j+3}.
\end{equation}
Similarly we can construct commuting higher charges for the glued model by using the charges of the original one,  and
thus we obtain a nearest neighbor 
integrable chain with local dimension $d^2$. In our arguments below we will switch back and forth between viewing the
spin chains and the charges in the original and the glued representations. This will gives us crucial clues about the
integrability structures.

First we look at the glued chain. It has a set of commuting local charges, thus we can assume that it is Yang-Baxter
integrable in the usual way. Thus there should
exist an auxiliary space $A$ and an $R$-matrix which generates the charges for this glued chain.
The auxiliary
space should have the same dimension as the physical spaces, which are now the tensor product spaces
$(j,j+1)$. Therefore we construct the auxiliary space $V_A$ as $V_a\otimes V_b$ where the two spaces labeled $a$ and $b$
are isomorphic to the physical spaces of the original chain. This way we obtain 
a commuting set of transfer matrices
\begin{equation}
  \label{eq:transfer3site}
   t(u)=\text{Tr}_A   R_{A,(L-1,L)}(u,0)\dots R_{A,(3,4)}(u,0)R_{A,(1,2)}(u,0),
\end{equation}
which generate the Hamiltonian and the charges of the glued chain.
Here we used the same letter for the $R$-matrix as before; the distinction between the different $R$-matrices is made by
denoting 
explicitly the indices of the spaces on which they act.

Above we set the inhomogeneity parameter of the $R$-matrix to $\xi_0=0$, which is just a choice for the zero point of the
rapidity parameters. Generally we can consider families of models by varying $\xi_0$, see for example
  \cite{Murakami-Hubbard-inhom} 
  for deformations of the Hubbard model. However, picking a particular model we are always free to set $\xi_0=0$ by a
  re-parametrization.

Since the glued chain is a nearest neighbor interacting model we can assume that the $R$-matrix is regular i.e.
\begin{equation}
   R_{(a,b),(j,j+1)}(0,0)=\Pe_{(a,b),(j,j+1)} = \Pe_{a,j} \Pe_{b,j+1}.
\end{equation}

Now let us return to the original spin chain. The transfer matrix constructed above can be understood as a non-local
operator acting on the original Hilbert space. The Taylor expansion of $t(u)$ on the glued chain gives the glued
charges, and if we view $t(u)$ as an operator acting on the original chain, then we see that it generates the charges of
the original model. This is a crucial observation. 

In the original spin chain all charges are translationally invariant, but the gluing procedure explicitly breaks this
invariance. The transfer matrix \eqref{eq:transfer3site} enjoys two-site invariance by construction:
\begin{equation}
 t(u)=\UU^{-2} t(u) \UU^2,
\end{equation}
where $\UU$ is the translation operator of the original chain, thus $\UU^2$ describes a single
site translation in the glued chain.
However, the charges of the original chain are one site invariant, and they are the Taylor coefficients of $t(u)$ if
viewed as an operator acting on the original chain.
This leads to the conclusion that the transfer matrix should also be translationally invariant:
\begin{equation}
   t(u)=\UU^{-1} t(u) \UU.
\end{equation}
This is a very strong condition. It implies that the two different gluing procedures (where we group sites $(2j,2j+1)$ or
$(2j+1,2j+2)$, respectively) should lead to the same global transfer matrix:
\begin{multline}
\label{Reltolt}
  \text{Tr}_{a,b} R_{(a,b),(L-1,L)}(u,0)\dots R_{(a,b),(1,2)}(u,0)=\\
   \text{Tr}_{a,b} R_{(a,b),(L,1)}(u,0)\dots R_{(a,b),(2,3)}(u,0).
\end{multline}
This relation motivates the existence of a proper Lax operator for the original chain, so that we could bypass the
gluing procedure. We expect that the only way that
\eqref{Reltolt} can be satisfied is if the corresponding $R$-matrix factorizes into a product of Lax operators. We
formulate the following:

\begin{conjecture}
  \label{conj:fact}
If the condition \eqref{Reltolt} holds in every even volume $L$ then the $R$-matrix of the glued
chain factorizes as
\begin{equation}
  \label{Rfact}
  R_{(a,b),(j,j+1)}(u,0)=\La_{a,b,j+1}(u)\La_{a,b,j}(u),
\end{equation}
where $\La_{a,b,j}(u)$ is a proper Lax operator for the original chain
which satisfies the following $RLL$ relation
\begin{multline}
  \label{eq:RLL}
R_{AB}(u,v)\La_{A,j}(u)\La_{B,j}(v)=\\=\La_{B,j}(v)\La_{A,j}(u) R_{AB}(u,v), 
\end{multline}
where $A$ and $B$ stand for pairs of auxiliary spaces, and $j$ labels a physical space.
\end{conjecture}

We did not find a proof for this Conjecture, but it is easy to see that the assumption \eqref{Rfact} naturally satisfies
\eqref{Reltolt}.

It is a simple consequence of Conjecture \ref{conj:fact}. that if the glued $R$-matrix is regular then the Lax operator
satisfies the three-site regularity condition  
\begin{equation}
\label{Linit}
  \La_{ab,j}(0)=\Pe_{a,j}\Pe_{b,j}.
\end{equation}
This follows simply from the observation 
  \begin{multline}
   R_{(1,2),(3,4)}(0,0) = \Pe_{2,4}P_{1,3} = 
   \Pe_{2,4} \Pe_{1,2} \Pe_{1,2} \Pe_{1,3}=\\
    =\left(\Pe_{1,4}\Pe_{2,4}\right) \left(\Pe_{1,3}\Pe_{2,3}\right)=
    \La_{1,2,4}(0)\La_{1,2,3}(0).
\label{eq:R0}
 \end{multline}
 
In general the RLL equations do not have unique solutions (up to normalization) if the R-matrix has gauge invariance 
\begin{equation}
 [ R_{A,B}(u,v),G_A(u)G_B(v)] = 0
\end{equation}
where $G_A(u)$ is a spectral parameter dependent $d\times d$ matrix. Indeed, assuming the gauge invariance the following transformed Lax operator
\begin{equation}
 G_A(u) \La_{A,j}(u)
\end{equation} 
is also a solution of the RLL equation \eqref{eq:RLL}. For the practical examples such symmetry of $R$-matrix is excluded and only global symmetry (spectral parameter independent symmetry) is allowed. In the following we concentrate on $R$-matrices with no gauge invariance.

Although we cannot prove Conjecture \ref{conj:fact}, we can prove the reverse statement:
\begin{theorem}
  \label{thm:fact}
   Taking a regular $R$-matrix $R_{A,B}(u)$ with no gauge invariance and a Lax operator $\La_{A,j}(u)$ which satisfy the $RLL$ relation \eqref{eq:RLL} and the regularity condition \eqref{Linit} then the $R$-matrix is factorized as \eqref{Rfact}.  
\end{theorem}
The proof is presented in Appendix \ref{sec:factproof}.

Let us now consider the transfer matrix of the original three-site interacting case.
We can now write it as
\begin{equation}
  \label{tL3}
  t(u)=\text{Tr}_{a,b}\La_{a,b,L}(u)\dots \La_{a,b,1}(u).
\end{equation}
This representation of the transfer matrix is manifestly translationally invariant; the formula already appeared in
\cite{sajat-cellaut}. 

The relation \eqref{Linit} leads to the initial condition
\begin{equation}
  \label{t0UU}
  t(0)=\UU^2.
\end{equation}
Let us now define the operator $\check \La_{a,b,j}(u)$ through
\begin{equation}
   \La_{a,b,j}(u)=\Pe_{a,j}\Pe_{b,j} \check \La_{a,b,j}(u).
\end{equation}
The condition \eqref{Linit} leads to
\begin{equation}
  \label{Lchinit}
  \check \La_{a,b,j}(0)=1.
\end{equation}
Computing the first logarithmic derivative of the transfer matrix as
\begin{equation}
  H=t^{-1}(0)\left. \partial_u t(u)\right|_{u=0}
\end{equation}
we obtain \eqref{H3} with
\begin{equation}
  \label{Lderiv}
  h_{j,j+1,j+2}=\left. \partial_u \check \La_{j,j+1,j+2}(u)\right|_{u=0}.
\end{equation}
Here we used
\begin{equation}
\label{ti0}
  t^{-1}(0)=\UU^{-2}.
\end{equation}

Summarizing these findings we formulate the following:
\begin{conjecture}
  \label{conj:3site}
Every integrable three-site Hamiltonian \eqref{H3}
has a three-site Lax operator in the form 
\begin{equation}
\La_{1,2,3}(u)= \Pe_{13} \Pe_{23}(1+uh_{123}+\mathcal{O}(u^{2})),
\end{equation}
and there exists an $R$-matrix for which the $RLL$-relation is satisfied
  \begin{multline}
R_{AB}(u,v)\La_{A,j}(u)\La_{B,j}(v)=\\=\La_{B,j}(v)\La_{A,j}(u) R_{AB}(u,v),
  \end{multline}
such that the $R$-matrix factorizes as \eqref{Rfact}. In the RLL relation above $A$ and $B$ stand for pairs of auxiliary spaces.
\end{conjecture}

Let us now also discuss two trivial solutions to the above relations, which bring us back to the nearest neighbor
chains.

We can choose
\begin{equation}
  \label{Labjtrivial}
   \La_{a,b,j}(u)=\La_{a,j}(u)\La_{b,j}(u),
 \end{equation}
 where $\La_{a,j}(u)$ is a Lax operator of a n.n. model. This choice satisfies all the requirements listed above. If we
 denote by $t^{(3)}(u)$ the transfer matrix constructed out 
of \eqref{Labjtrivial} using \eqref{tL3} and by $t^{(2)}(u)$ the simple transfer matrix constructed from $\La_{a,j}(u)$
according to \eqref{teredeti}
then we obtain the relation
\begin{equation}
  t^{(3)}(u)=\left(t^{(2)}(u)\right)^2.
\end{equation}
This implies that from $t^{(3)}(u)$ we would obtain the same nearest neighbor Hamiltonian as from $t^{(2)}(u)$, but
multiplied with factor of 2. 

An alternative trivial choice is
\begin{equation}
  \label{Labjtriv2}
  \La_{a,b,c}(u)= \Pe_{a,b} \La_{a,c}(u)
\end{equation}
It can be seen that this leads to two decoupled spin chains that are placed on the odd and even sub-lattices of the
original spin chain, such that we have nearest neighbor interactions within each sub-lattice.

These trivial examples bring us to an important conclusion: {\it the real source of the three site interaction is a
  coupling of the auxiliary 
spaces which can not be factorized as in \eqref{Labjtrivial} or as in \eqref{Labjtriv2}}.

\subsection{Conserved charges in the three-site interacting case}

\label{sec:Q5}

Let us now consider the higher conserved charges in the three-site interacting models, which are computed from the
higher logarithmic derivatives of the transfer matrix. The next charge is computed from
\begin{equation}
  (\partial_u)^2 \log(t(u))=t^{-1}(u)t''(u)-(t^{-1}(u)t'(u))^2
\end{equation}
after taking the derivatives and eventually substituting $u=0$. Taking into account the definition \eqref{tL3}, the initial
conditions \eqref{Linit} and also the inverse \eqref{ti0} we obtain a 5 site operator, which can be written as
\begin{multline}
\label{Q5a}
  Q_5=\sum_j [h_{j,j+1,j+2},h_{j+1,j+2,j+3}+h_{j+2,j+3,j+4}]-\\ 
- (h_{j,j+1,j+2})^2+ \check \La_{j,j+1,j+2}''(0).
\end{multline}
This is a generalization of formula \eqref{Q3form1} to the three site interacting case. We see that the density of $Q_5$
depends on a term which is completely determined by the Hamiltonian density, but it also includes an additional
three-site operator, which was included in the equation \eqref{q5d} that we announced earlier.

The simplest cases are those when the following inversion relation holds:
\begin{equation}
\label{L3inv}
   \check \La_{a,b,j}(u) \check \La_{a,b,j}(-u)=1.
\end{equation}
Taking second derivative in $u$ and substituting $u=0$ we obtain that the last two terms in \eqref{Q5a} cancel.

At present it is not clear whether all three-site interacting models satisfy \eqref{L3inv}. In our classification we
found that we can always choose conventions such that \eqref{L3inv} holds. In the case of the $U(1)$-invariant models we
actually allowed for an additional three-site 
operator in  the density of $Q_5$, performed the classification using the condition $[H,Q_5]=0$ and did not find
additional models. However, this might be a peculiarity of the $U(1)$-invariant models.

We also note a gauge freedom which does actually change the representation of the charge $Q_5$. Considering a Hamiltonian
density $h_{1,2,3}$ we can construct a new one by
\begin{equation}
  \label{h123v}
  h'_{1,2,3}=h_{1,2,3}+g_{1,2}-g_{2,3},
\end{equation}
where $g_{1,2}$ is an arbitrary two-site operator. Clearly, the density $h'_{1,2,3}$ leads to the same global
Hamiltonian, because the additional terms add up telescopically to zero. However, the commutator in \eqref{Q5a}
will be a different operator. The dependence on $g_{1,2}$ does not drop 
out, whereas the global $Q_5$ charge can not change. This paradox is resolved by noting that the three-site
operator in the second line of \eqref{Q5a} also changes, because the Lax operator needs to be modified so that its first
derivative can reproduce \eqref{h123v}. Altogether this modification cancels the additional terms that appear from the
commutator in \eqref{Q5a}, so that $Q_5$ does not change as expected. In our concrete
examples we always found a gauge where \eqref{L3inv} holds.

Let us now also consider the higher charges. Taking further derivatives we see that the range of the charges increases
by two sites after every new derivative. This 
means that the allowed indices $\alpha$ for the charges $Q_\alpha$ are $3,5,7,\dots$ with $H=Q_3$. This is clearly
different from the n.n. chains where typically all integers are allowed.

In specific cases it can happen that there is a one-site or a two-site charge commuting with the transfer matrix
above; examples will be shown below.
In all such cases we found that the smaller charges are not dynamical. If they are non-trivial then the model has to be 
nearest neighbor interacting.

\subsection{Partial classification for three site interacting spin-1/2 chains}

\label{sec:classification}

Our method is to make an Ansatz for $h_{j,j+1,j+2}$ and $\tilde
h_{j,j+1,j+2}$, to compute 
$Q_5$ through \eqref{q5d}, and finally to enforce the commutativity $[H,Q_5]=0$. We find that this relation indeed gives
us a number of interesting new models. Afterwards we also look for the corresponding Lax operators and $R$-matrices. The
idea for finding the Lax operator is rather simple: We are looking for a one-parameter family of three-site operators,
within the specified Ansatz, and we enforce that the transfer matrix \eqref{tL3} commutes with the
Hamiltonian. Afterwards we also check the initial condition \eqref{Linit} and the first derivative according to
\eqref{Lderiv}. Finally the $R$-matrices can be found simply from \eqref{RLL} which becomes a linear equation for
them. The Yang-Baxter relation for the $R$-matrices can be checked afterwards.

It is important for the classification to exclude ``trivial'' solutions which would not lead to new physical
behaviour. Such trivial cases include simply taking a nearest neighbor Hamiltonian density $h^{(nn)}_{j,j+1}$ and
choosing either $h_{j,j+1,j+2}=h^{(nn)}_{j,j+1}$ or $h_{j,j+1,j+2}=h^{(nn)}_{j,j+2}$. The first choice simply just gives
back the original two-site model, whereas the second choice gives two decoupled nearest neighbor chains living on the
odd and even sub-lattices of the new model. We encountered both cases among the results of the classification, but we
will not include them in the lists to be presented below.

Curiously we found that in some restricted parameter spaces $\tilde h_{j,j+1,j+2}=0$, for example this holds for all
$U(1)$-invariant models. However, at present we can not exclude models in other classes with a non-zero $\tilde
h_{j,j+1,j+2}$. 

We performed the classification in three specific cases.

\subsubsection{$SU(2)$ invariant models}

It is relatively easy to impose $SU(2)$ invariance on the three-site Hamiltonian density $h_{j,j+1,j+2}$: we require that
it has to be built as a sum of permutation operators that act on the three sites $j,j+1,j+2$. We do not impose space
reflection invariance. With these conditions we found {\bf no non-trivial three site models}. Trivial solutions include
$h_{j,j+1,j+2}=\Pe_{j,j+1}$ which describes the Heisenberg spin chain, $h_{j,j+1,j+2}=\Pe_{j,j+2}$ which describes two
independent Heisenberg chains on two sub-lattices, and $h_{j,j+1,j+2}=q_3(j)$ with $q_3(j)$ being the density of the
third charge in the Heisenberg chain.

\bigskip

\subsubsection{$U(1)$ invariant models with space reflection invariance}

We investigated models where the Hamiltonian commutes with the charge $Q_1=S_z$, which generates a global $U(1)$
group. We also required space reflection invariance, but allowing for a gauge freedom of the type \eqref{h123v}.

We found two families of models.

{\bf The Bariev model.} It is given by the Hamiltonian
\begin{equation}
  \label{Hbariev}
  H=\sum_j
   \left[\sigma^-_j\sigma^+_{j+2}+\sigma^+_j\sigma^-_{j+2}\right] \frac{1-U\sigma^z_{j+1}}{2},
\end{equation}
where $U$ is a coupling constant. The model was first proposed in \cite{bariev-model} as a zig-zag spin ladder. Special
points of the model are $U=\pm 1$
where it becomes identical to the folded XXZ model in the bond picture \cite{folded1,folded2,sajat-folded}. We
investigated the fundamental Lax and $R$-matrices of the model given in \cite{bariev-lax-1} and found that they satisfy
the equations derived above, including the initial condition \eqref{Linit} and the factorization \eqref{Rfact} (for the
Lax operators see also \cite{bariev-lax-2,bariev-lax-3}). To
obtain the desired formulas we just applied certain permutations on the basis vectors so that
 \eqref{Linit} and \eqref{Rfact} would hold in the form given above. 

{\bf The hard rod deformed XXZ model.} It is given by
  \begin{multline}
  \label{Hhrdef2}
   H=\sum_j  \left[\sigma^-_jP^\bullet_{j+1}\sigma^+_{j+2}+\sigma^+_jP^\bullet_{j+1}\sigma^-_{j+2}\right.\\
\left.    -\Delta (P^\circ_j P^\bullet_{j+1}P^\bullet_{j+2}+P^\bullet_j P^\bullet_{j+1}P^\circ_{j+2})\right].
\end{multline}
Here $\Delta$ is a coupling constant. Up to our best knowledge this is a new model, but it is closely related to the
so-called constrained XXZ model treated earlier in
\cite{constrained1,constrained2,constrained3,pronko-abarenkova-ising-limit,xxz-triple-point}. The complete solution of
the new model and the discussion of
its special physical properties will be discussed in an upcoming publication. We note that at the special point
$\Delta=0$ this model also becomes equal to the folded XXZ model in the bond picture. Thus these two families of models
intersect at the points $\Delta=0$ and $U=1$.

\subsubsection{Hamiltonians of the IRF type}

\label{sec:IRFH0}

A further special class of models are those when the three-site operator
$h_{j,j+1,j+2}$ acts diagonally on the first and the last sites. This means that the spins at site $j$ and $j+2$ can be
considered as control bits that influence the action on the site $j+1$. The most general form for such an operator is
\begin{equation}
  \label{IRFh}
  h_{j,j+1,j+2}=\sum_{a,b=0,1}  P^a_j h^{ab}_{j+1} P^b_{j+2},
\end{equation}
where now $h^{ab}$ stands for a collection of four Hermitian matrices corresponding to the indices $a,b=0,1$.

The search for such Hamiltonians is motivated by recent studies on quantum gates and cellular automata, which we
discussed in Section \ref{sec:cellintro}.
 In particular, such Hamiltonians can be considered as continuous time version of the IRF models treated in
 \cite{prosen-cellaut}. 

We treat these models separately in Section \ref{sec:IRFH}.

\section{Integrable quantum circuits}

\label{sec:circuit}

In this Section we present our results for the brickwork type quantum circuits. We introduce a number of closely related
constructions, with different types of integrability properties.

Let us recall the general formulas for the brickwork circuits as explained in \ref{sec:qgatesintro}. We are building 
Floquet-type cycles with time period $\tau$:
\begin{equation}
  \label{floquetb}
  \mathcal{V}=\mathcal{V}_\tau\dots\mathcal{V}_1
\end{equation}
with the update steps being
\begin{equation}
  \label{Vjb}
  \mathcal{V}_l=\prod_{k}  U^{(\ell)}(x_k+\Delta_l).
\end{equation}
Here $x_k$ are coordinates for the unitaries and
$\Delta_l$ are the displacements that distinguish the different single step updates within the Floquet cycle.

\subsection{Integrable Trotterization for nearest neighbor interacting chains}

Here we review the construction of \cite{integrable-trotterization}, which can be applied for integrable nearest
neighbor chains; the key ideas go back to the light cone regularization of the Quantum Field Theories, see for example
\cite{DdV0,Volkov,Faddeev-Volkov}. 

\begin{figure}
\centering
\includegraphics[width=0.6\columnwidth]{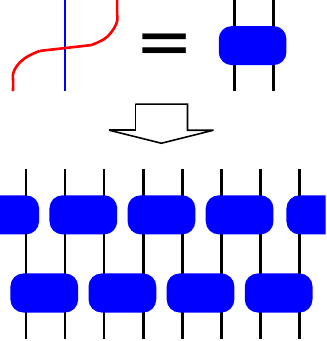}
\caption{Graphical illustration of the period-2 Floquet cycle $( t(\nu))^{-1}t(\mu)$.}
\label{fig:2siteQgates}
\end{figure}

We build a brickwork quantum circuit using two-site unitaries ($\ell=2$) with Floquet period $\tau=2$. Correspondingly, the
coordinates for the unitaries are $x_k=2k$ and the shift is $\Delta_l=l$, see Figure \ref{fig:2siteQgates}.

The starting point is the
$R$-matrix $R(u,v)$ which is supposed to be a regular solution of the YB equations \eqref{YB}. We do not assume that the
$R$-matrix is of difference form.

\begin{figure}
\centering
\includegraphics[width=0.7\columnwidth]{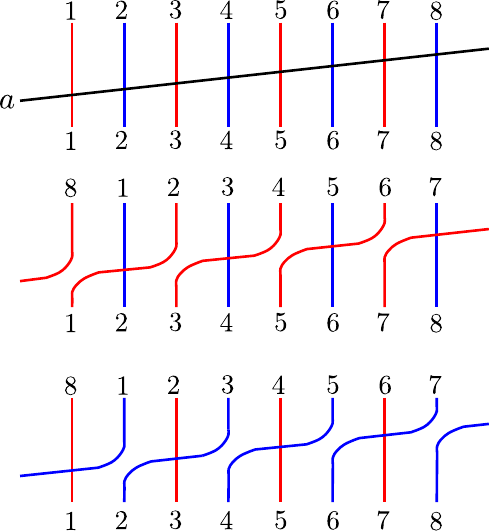}
\caption{Graphical illustration of transfer matrices $t(u)$, $t(\mu)$ and $t(\nu)$. The corresponding spectral
  parameters of black, red and blue lines are $u$, $\mu$ and $\nu$, respectively. The line avoidings are a result of the
regularity condition \eqref{eq:regular}.}
\label{fig:trotter}
\end{figure}

On a chain of length $L=2k$ we build an {\it inhomogeneous} transfer matrix with alternating inhomogeneities $\mu,\nu$ as
\begin{multline}
  \label{inhom}
  t(u)=\text{Tr}_a \Big[R_{a,L}(u,\nu)R_{a,L-1}(u,\mu) \dots \\
  \dots R_{a,2}(u,\nu)R_{a,1}(u,\mu)\Big].
\end{multline}
Using the regularity condition \eqref{eq:regular} we obtain two special points for the transfer matrix, where it becomes
a product of distinct quantum gates multiplied by an overall translation (see figure \ref{fig:trotter})
\begin{align}
  t(\mu) &= \UU \check R_{L-1,L}(\mu,\nu)   \dots \check R_{1,2}(\mu,\nu),\\
  t(\nu) &= \UU \check R_{L,1}(\nu,\mu)   \dots \check R_{2,3}(\nu,\mu).
\end{align}
Let us assume that the inversion relation \eqref{Rinv} holds without scalar factors
\begin{equation}
  \label{Rchinv}
 \check R_{1,2}(u,v) \check R_{1,2}(v,u) = 1.
\end{equation}
Then we obtain
\begin{equation}
  ( t(\nu))^{-1}= \check R_{L,1}(\mu,\nu)   \dots \check R_{2,3}(\mu,\nu) \UU^{-1}.
\end{equation}
Finally we see that the operator product
\begin{equation}
   ( t(\nu))^{-1} t(\mu)
\end{equation}
can be interpreted as a period-2 Floquet cycle of the form \eqref{floquet} with a two-site gate
\begin{equation}
  U^{(2)}(j)=\check R_{j,j+1}(\mu,\nu),
\end{equation}
see figure \ref{fig:2siteQgates}.

The requirement of unitarity puts constraints on the $R$-matrix and the spectral parameters $\mu,\nu$. However, in the
typical cases there are simple choices which fulfill unitarity, which depend on the real analyticity of the
$R$-matrix. It follows from \eqref{Rchinv} that the $R$-matrix is unitary for a pair $\mu,\nu$ if
\begin{equation}
 \left( \check R_{1,2}(\mu,\nu)\right)^\dagger= \check R_{1,2}(\nu,\mu).
\end{equation}
In the simple case of the XXZ spin chain with $\Delta=\cosh(\eta)$ we can choose the following representation of the $R$-matrix:
\begin{equation}
  \label{XXZRcheck}
\check R(\mu,\nu)=
  \begin{pmatrix}
    1 & & & \\
    & c(\mu-\nu)  & b(\mu-\nu) & \\
    & b(\mu-\nu) & c(\mu-\nu) & \\
        & & & 1\\
  \end{pmatrix}
\end{equation}
with
\begin{equation}
  c(u)=\frac{\sinh(\eta)}{\sinh(u+\eta)},\qquad b(u)=\frac{\sinh(u)}{\sinh(u+\eta)}.
\end{equation}
It can be seen that $\check R(\mu,\nu)$ is unitary if either $\eta$ is real ($\Delta>1$) and $\mu-\nu$ is purely
imaginary, or if $\eta$ is purely imaginary ($\Delta<1$) and $\mu-\nu$ is real.

\subsection{Integrable quantum circuits for three-site models -- Construction 1}

\label{sec:threegates}

Now we construct quantum circuits for the three site interacting models, which
can be applied for every model which fits into our framework laid out in Section \ref{sec:int}. The specific case of the
IRF type models is treated later in \ref{sec:IRFgates}, where we present a different type of quantum circuit adapted
to the special form of those Lax operators. 

We present two different Floquet circles for the three site models. In the first case we
generalize the 
results from the previous Subsection to the present case: this is a rather straightforward construction from a technical
point of view, but it leads to brickwork circuits with ``untouched'' sites.
An alternative construction is presented in the next Subsection. That one leads to a tightly packed brickwork circuit,
but its integrability structure is more involved.

In the first case we use the glued chain, where we group together pairs of sites of the
original chain, see the 
derivations in Section \ref{sec:threesite}. Then the idea is to generalize the definition \eqref{inhom} to the present
case with the grouped sites. This way we obtain a quantum circuit with 4-site unitaries, because the $R$-matrix of
the glued chain acts on pairs of sites. However, at special points we can make use of the factorization \eqref{Rfact} in
order to obtain the three site unitaries. 

We take a chain of length $L=4k$ and construct an inhomogeneous transfer matrix for the grouped sites with alternating
inhomogeneities $\mu,\nu$:
\begin{multline}
  t(u)=\text{Tr}_{ab} \big[R_{(a,b)(4k-1,4k)}(u,\nu)\dots 
  \\ \dots R_{(a,b)(3,4)}(u,\nu)R_{(a,b)(1,2)}(u,\mu)\big]
\end{multline}
Here $a,b$ stand for the two auxiliary spaces of the model.
This family of transfer matrices is commuting.
Special points are $u=\mu,\nu$ with
\begin{equation}
  \begin{split}
    t(\mu)&=\UU^2  \check R_{(4k-3,4k-2)(4k-1,4k)}(\mu,\nu) \dots \check R_{(1,2)(3,4)}(\mu,\nu),\\
     t(\nu)&=\UU^2  \check R_{(4k-1,4k)(1,2)}(\nu,\mu) \dots \check R_{(3,4)(5,6)}(\nu,\mu).\\   
  \end{split}
\end{equation}
Note the appearance of $\UU^2$, the translation operator by two sites.

Similar to the nearest neighbor case, we take now the inverse of $t(\mu)$, which becomes
\begin{equation}
     t^{-1}(\nu)=  \check R_{(4k-1,4k)(1,2)}(\mu,\nu) \dots \check R_{(3,4)(5,6)}(\mu,\nu) \UU^{-2}.
\end{equation}
Finally we define the Floquet cycle as
\begin{multline}
  \label{4FV}
     \mathcal{V}=t^{-1}(\nu) t(\mu)=\\
     =\check R_{(4k-1,4k)(1,2)}(\mu,\nu) \dots \check R_{(3,4)(5,6)}(\mu,\nu) \times\\
  \times   \check R_{(4k-3,4k-2)(4k,4k-1)}(\mu,\nu) \dots \check R_{(1,2)(3,4)}(\mu,\nu).
   \end{multline}
This can be understood as a cycle with period $\tau=2$ and with the four site gates
\begin{equation}
     U^{(4)}(j)=\check R_{(j,j+1)(j+2,j+3)}(\mu,\nu).
\end{equation}
The coordinates of the gates are $x_k=4k$ and the displacements are $\Delta_l=2l$ with $l=1,2$. Similar to the nearest
neighbor cases we need to impose restrictions on $\mu,\nu$ to obtain a gate which is unitary. This restriction depends
on the model.

\begin{figure}
\centering
\includegraphics[width=0.8\columnwidth]{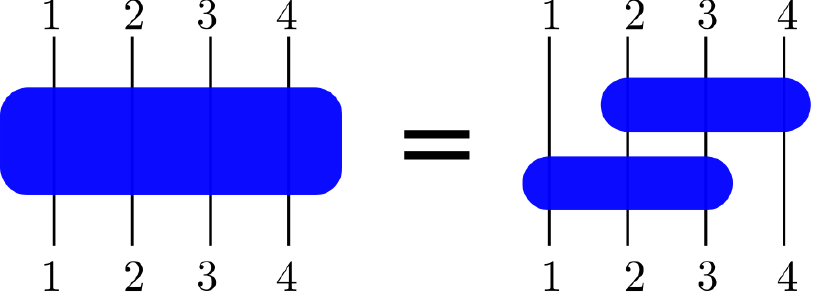}
\caption{Graphical illustration of eq. \eqref{eq:RLch}.}
\label{fig:RLch}
\end{figure}

\begin{figure}
\centering
\includegraphics[width=0.99\columnwidth]{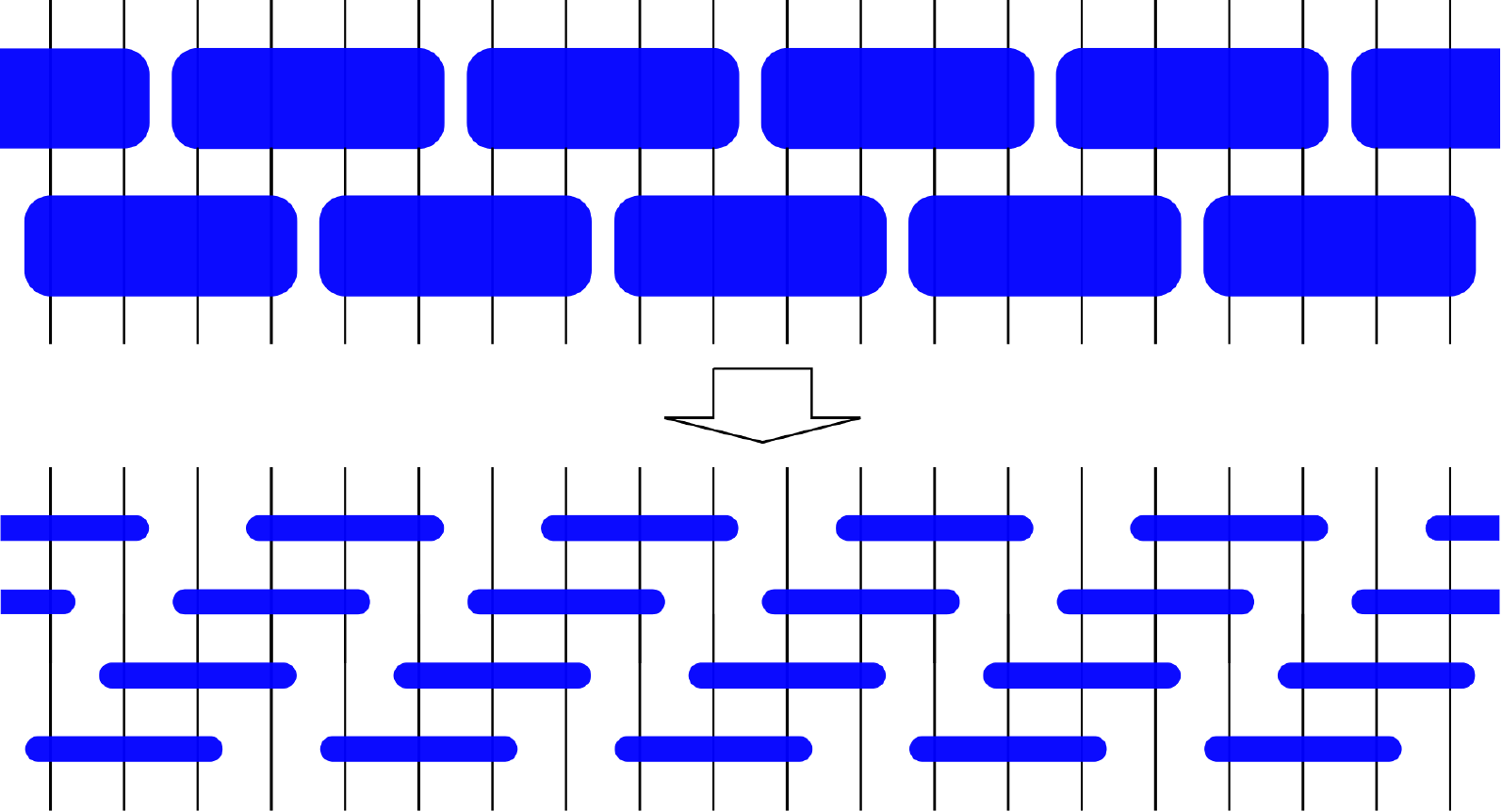}
\caption{In the top of the figure we can see the period 2 Floquet cycle for the glued chain. We changed the
  $R$-matrices from Fig. \ref{fig:2siteQgates} to $\check R_{(12),(34)}(\mu,\nu)$. In the bottom we substitute $\nu=0$
  and use the factorization 
  property \eqref{eq:RLch}. Thus we get a period 4 Floquet cycle with 3-site gates so that every fourth spin is
  untouched at every step. }
\label{fig:3siteQgates}
\end{figure}

A quantum circuit with three site gates is obtained by substituting $\nu=0$. Then we use the factorization
\begin{equation} \label{eq:RLch}
\check R_{(j,j+1),(j+2,j+3)}(\mu,0) = 
\check \La_{j+1,j+2,j+3}(\mu)\check \La_{j,j+1,j+2}(\mu).
\end{equation}
which follows from \eqref{Rfact}. For a graphical interpretation of this relation see Figure \ref{fig:RLch}.
\begin{figure}
\centering
\includegraphics[width=0.8\columnwidth]{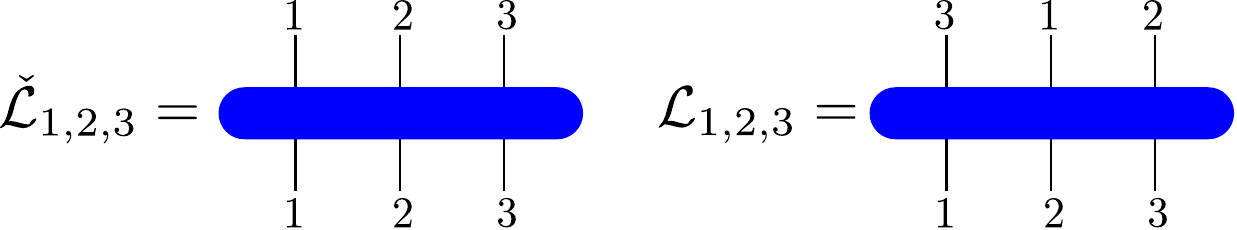}
\caption{Graphical illustration of operators $\check \La_{123}$ and $\La_{123}$.}
\label{fig:Lconv}
\end{figure}
Substituting this factorization into \eqref{4FV} we obtain a Floquet cycle with period $\tau=4$, three-site gates
\begin{equation}
  \label{U3La}
  U^{(3)}(j)=\check \La_{j,j+1,j+2}(\mu),
\end{equation}
coordinates $x_k=4k$ and displacements $\Delta_l=l$ with $l=1,2,3,4$. For a graphical interpretation of this quantum
circuit see Figure \ref{fig:3siteQgates}.

Disadvantages of this construction are that the
quantum circuit does not have left-right symmetry, and that the three-site unitaries are not tightly packed: every fourth
spin is left untouched at every time step.

\subsection{Integrable quantum circuits for three-site models -- Construction 2}

\begin{figure}
\centering
\includegraphics[width=0.99\columnwidth]{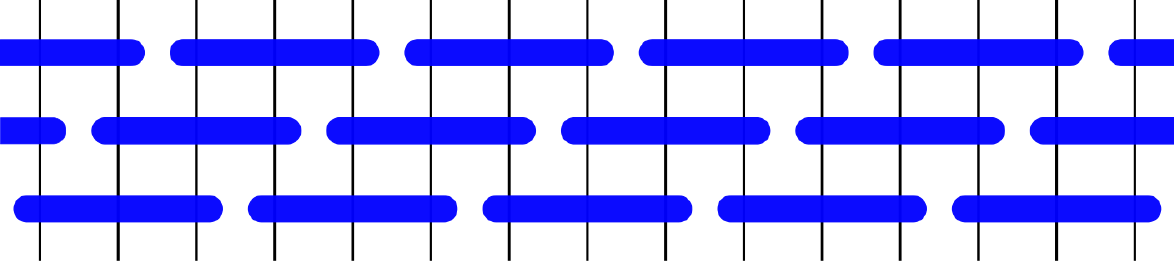}
\caption{A tightly packed quantum circuit with three site gates. This network has periodicity 3 in both the space and
  time directions. Accordingly, we can prove the existence of commuting transfer matrices built from three rows.}
\label{fig:3siteDiag2Diag}
\end{figure}

Now we also present a quantum circuit which is tightly packed. 
We take three site gates defined by \eqref{U3La} and construct a circuit with period $\tau=3$, coordinates $x_k=3k$ and displacements
$\Delta_l=l$ with $l=1,2,3$. For a graphical interpretation see Fig. \ref{fig:3siteDiag2Diag}.

\begin{figure}
\centering
\includegraphics[width=0.39\columnwidth]{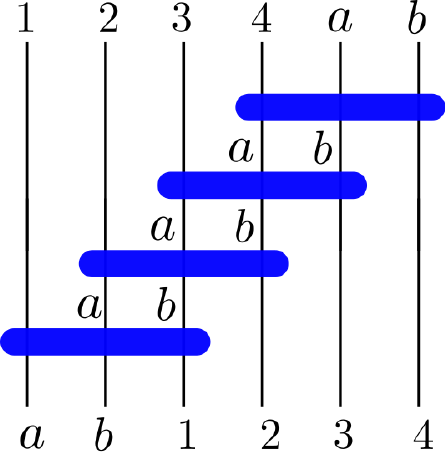}
\caption{Graphical interpretation of diagonal to diagonal transfer matrices, which are identical to those defined in
  \eqref{tL3}, but drawn simply in a different arrangement, compare also with Fig. \ref{fig:transfer}. This transfer
  matrix can also be used to build the circuit shown in \ref{fig:3siteDiag2Diag}, this was first worked out in
  \cite{sajat-cellaut}.
 }
\label{fig:diag2diag}
\end{figure}

Such a circuit was already introduced in \cite{sajat-cellaut}, however, the integrability was only shown for the
diagonal-to-diagonal transfer matrices, which are identical to the ones defined by \eqref{tL3} (for a pictorial
interpretation see Fig. \ref{fig:diag2diag}). Now we also develop the
commuting row-to-row transfer matrices for this quantum circuit.

We start with the three site Lax operator $\check \La(u)$ and the associated $R$-matrix $\check R_{ab,(j,j+1)}(u,v)$.
We construct an operator acting on 5 spaces
\begin{align}
  \check \Ra_{1,2,3,4,5}(\theta,u) 
  &= \check  \La^{-1}_{1,2,3}(u) \check R_{2,3,4,5}(\theta,u) \check \La_{1,2,3}(\theta) \nonumber\\
&=\check  \La_{3,4,5}(\theta) \check R_{1,2,3,4}(\theta,u) \check \La^{-1}_{3,4,5}(u).
\end{align}
For the transfer matrices we will also need an operator which acts on one more space with a permutation, therefore we define
\begin{equation}
  \label{R6}
  \Ra_{(1,2,3),(4,5,6)}(\theta,u)= \Pe_{1,4} \Pe_{2,5} \Pe_{3,6}
  \check \Ra_{1,2,3,4,5}(\theta,u).
\end{equation} 
This matrix satisfies the YB equation
\begin{multline}
  \label{YB6}
 \Ra_{(1,2,3),A}(\theta,u) \Ra_{(1,2,3),B}(\theta,v) \Ra_{A,B}(u,v) = \\
 \Ra_{A,B}(u,v) \Ra_{(1,2,3),B}(\theta,v) \Ra_{(1,2,3),A}(\theta,u),
\end{multline}
where $A=(4,5,6)$ and $B=(7,8,9)$ stand for additional two triplets of auxiliary spaces. Relation \eqref{YB6} can be
checked by direct substitution of the definitions above, and making use of the RLL relations \eqref{RLL}
and YB relations \eqref{YB} applied to the $R$-matrix in question.

Then we a construct a transfer matrix for our chain after grouping together triplets of physical spaces.
The precise formula with auxiliary space $A=(a,b,c)$ reads
\begin{multline}
  \label{eq:IRFtransfer5}
 t(u) =
 \mathrm{tr}_{A} \bigl[
 \Ra_{(1,2,3),A}(\theta,u) \Ra_{(4,5,6),A}(\theta,u) \dots \\
 \Ra_{(L-5,L-4,L-3),A}(\theta,u) \Ra_{(L-2,L-1,L),A}(\theta,u) \bigr].
\end{multline}
Note that the relative ordering of the physical triplets is such that we proceed backwards on the chain, as in
\eqref{eq:transferRev}.
  This is due to certain technical details, so that in the end we can
obtain the desired quantum circuit.

\begin{figure}
\centering
\includegraphics[width=0.7\columnwidth]{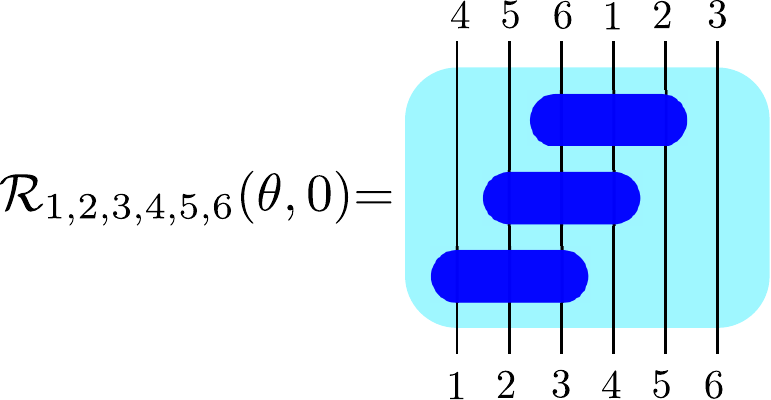}
\caption{Graphical representation for the factorization of $\Ra_{1,2,3,4,5,6}(\theta,0)$ into the three site gates given
  by $\La(\theta)$. Note that the last vector space is also involved in the permutations, which is
  necessary in order to build the desired transfer matrix.}
\label{fig:RR}
\end{figure}

\begin{figure}
\centering
\includegraphics[width=0.99\columnwidth]{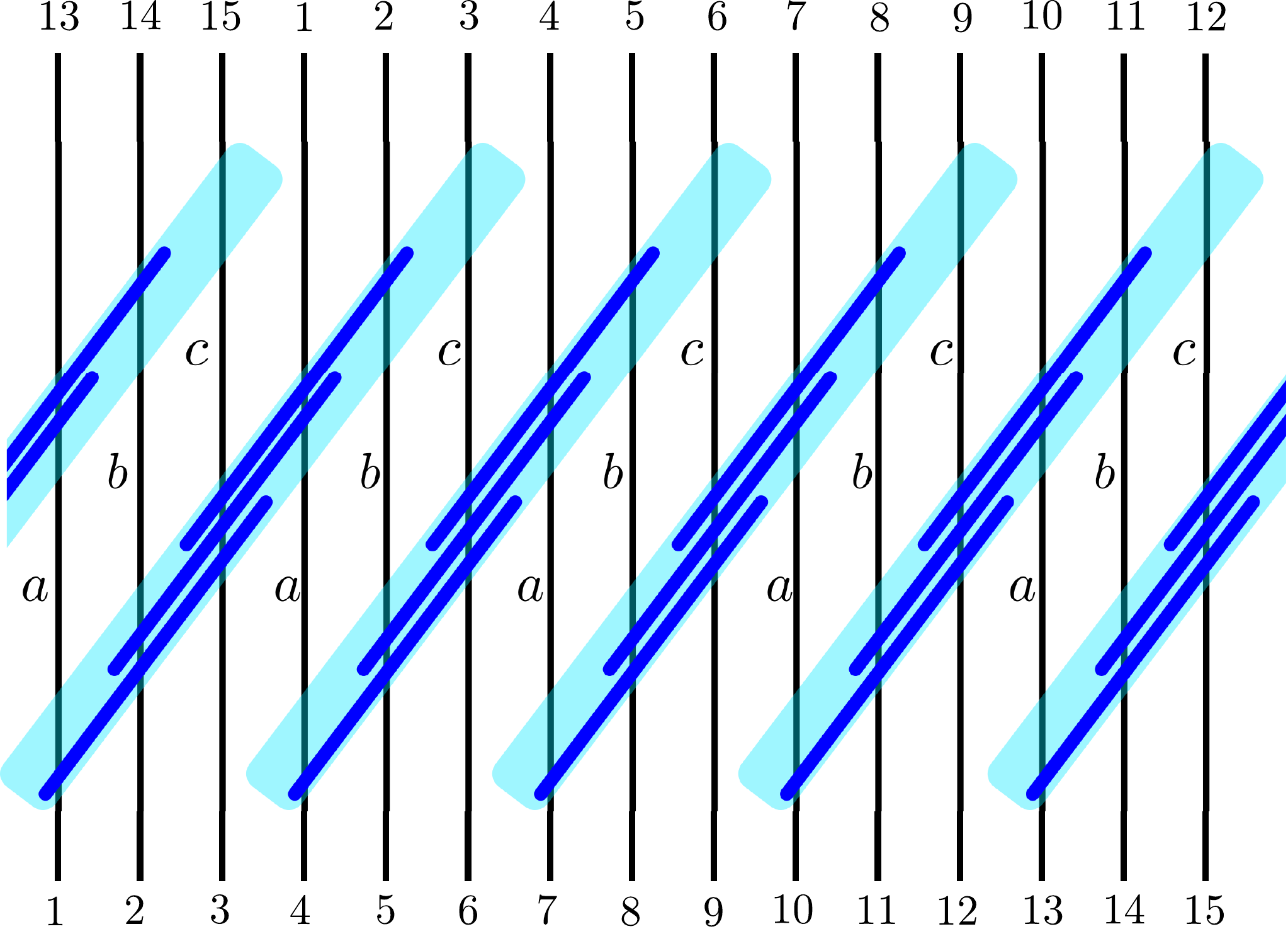}
\caption{Graphical proof of eq. \eqref{eq:update3site}. The labels $(a,b,c)$ stand for the triplet of auxiliary spaces which are
  present in the definition \eqref{eq:IRFtransfer5}. The big lightly shaded rectangles stand for the action of the $\Ra$
operators. Within each big rectangle we denoted the factorization into the product of $\La$ operators, as shown in
Fig. \ref{fig:RR}. Then the quantum circuit of Fig. \ref{fig:3siteDiag2Diag} is obtained by ``straightening out'' the
angles of the three site gates, noticing that they perfectly fit together to build the tightly packed circuit. Notice
that this circuit also inlcudes a translation by 3 sites, which is canceled by $t^{-1}(\theta)=\UU^{3}$ in \eqref{eq:update3site}.} 
\label{fig:3siteSzoros}
\end{figure}

These transfer matrices commute, if we regard $\theta$ as a fixed parameter.
The special point $u=\theta$ gives the translation operator by three sites:
\begin{equation}
 t(\theta) = \mathcal{U}^{-3}.
\end{equation}
The other special point is obtained by setting $u=0$. In this case we get
\begin{equation}
  \label{eq:update3site}
 t(\theta)^{-1} t(0) = \mathcal{V}_3\mathcal{V}_2\mathcal{V}_1.
\end{equation}
where the equal time update steps $\mathcal{V}_j$ are defined by
\begin{equation}
  \mathcal{V}_j=\prod_{k}  U^{(3)}(3k+j), \quad U^{(3}(j)=\check \La_{j,j+1,j+2}(\theta).
\end{equation}
For the proof of the initial condition \eqref{eq:update3site} we need to use the factorization condition \eqref{Rfact}
which eventually leads to
\begin{equation}
 \check \Ra_{1,2,3,4,5}(\theta,0)  = 
 \check \La_{3,4,5}(\theta) \check \La_{2,3,4}(\theta)   \check \La_{1,2,3}(\theta),
\end{equation}
where we also used the initial condition $\check \La_{1,2,3}(0)=1$. A graphical interpretation of this equation (applied
for the six-site object $\Ra_{1,2,3,4,5,6}$) is given in Fig 
\ref{fig:RR}. Multiplying the operators thus obtained, together with the permutations introduced in \eqref{R6} will
eventually lead to \eqref{eq:update3site}. A graphical proof is shown on Fig. \ref{fig:3siteSzoros}.

\section{Interaction-round-a-face type models}

\label{sec:IRFsec}

In this Section we study the IRF type models. Our motivations come from recent results in the literature about
elementary cellular automata, see the  discussion in Section \ref{sec:cellintro}. It is our goal to embed these models
into the algebraic framework of medium 
range models. We will see that the special structure of these models leads to unique constructions for the quantum
circuits, and also to the discovery of new algebraic structures that are connected to the ``face weight formulation'' of
the Yang-Baxter relation.

We are looking for three site integrable models where the Lax operator has the special form
\begin{equation}
  \label{IRFLax}
  \check \La_{j,j+1,j+2}(u)=\sum_{a,b=0,1} P_j^a f^{ab}_{j+1}(u)  P^b_{j+2},
\end{equation}
where $f^{ab}(u)$ is a collection of four $u$-dependent matrices. Such Lax operators will be used to build quantum
circuits: they will play the role of the three-site unitaries $U^{(3)}(j)$. The assumption \eqref{IRFLax} leads to
unique algebraic constructions and quantum circuits, which 
would be meaningless without the special structure of the Lax operator.

Later in Subsection \ref{sec:IRFH} we
perform a classification of such models, and in \ref{elementarycellaut} we also discuss the concrete cases that
eventually lead to the elementary 
cellular automata. 

\subsection{Integrability structures and quantum circuits}

\label{sec:IRFgates}

In general we have $R$- and Lax-matrices which satisfy the $RLL$ relation which we now write in the form
  \begin{multline}
      \label{RLLch}
 \check{R}_{23,45}(u,v)\check{\La}_{123}(u)\check{\La}_{345}(v)=\\=\check{\La}_{123}(v)\check{\La}_{345}(u)\check{R}_{12,34}(u,v).
  \end{multline}
It follows from \eqref{IRFLax} that now the Lax-operators satisfy an extra condition
\begin{equation}
  \label{IRFLcomm}
 [\check \La_{123}(u), \check \La_{345}(v)] = 0.
\end{equation}
Using this requirement the $RLL$ relation reads as
\begin{multline}
\check{\La}_{345}(u)^{-1}\check{R}_{23,45}(u,v)\check{\La}_{345}(v)= \\
\check{\La}_{123}(v)\check{R}_{12,34}(u,v)\check{\La}_{123}(u)^{-1}.
\end{multline}
We can see that the l.h.s. and the r.h.s. act trivially on the spaces $1$ and $5$ respectively, therefore they have to
be equal to
a three-site operator which we denote as $\check \Ga_{234}$. Thus we find
\begin{align}
 \check \Ga_{234}(u,v) &= \check{\La}_{345}(u)^{-1}
 \check{R}_{23,45}(u,v) \check{\La}_{345}(v), \label{eq:Gdef}\\
 \check \Ga_{234}(u,v) &= \check{\La}_{123}(v)
 \check{R}_{12,34}(u,v) \check{\La}_{123}(u)^{-1}. \label{eq:Gdef2}
\end{align}
We can express the $R$-matrices with $\Ga$ and $\La$ in two ways:
\begin{align}
 \check{R}_{12,34}(u,v)	&= \check{\La}_{123}(v)^{-1}
 \check{\Ga}_{234}(u,v) \check{\La}_{123}(u),\label{eq:RG2}\\
 \check{R}_{23,45}(u,v)	&= \check{\La}_{345}(u)
 \check{\Ga}_{234}(u,v) \check{\La}_{345}(v)^{-1},
\end{align}
or equivalently
\begin{equation} \label{eq:RG1}
 \check{R}_{12,34}(u,v) = \check{\La}_{234}(u)
 \check{\Ga}_{123}(u,v) \check{\La}_{234}(v)^{-1}.
\end{equation}
The consistency of the two expressions for the $R$-matrix leads to the equation
\begin{multline}
 \label{eq:GLL}
\check{\Ga}_{234}(u,v)\check{\La}_{123}(u)\check{\La}_{234}(v)=\\=\check{\La}_{123}(v)\check{\La}_{234}(u)\check{\Ga}_{123}(u,v).  
\end{multline}
This relation is similar to eq. \eqref{RLLch}, but there are important differences: Here the supports for the Lax
operators overlap at 
two spaces at both the l.h.s. and the r.h.s., whereas in \eqref{RLLch} they overlap only at a single space. We call this
equation the GLL 
relation. Below we show that it is equivalent to the ``face weight'' formulation of the RLL relation.

We can easily calculate this $G$-operator at special values of spectral parameters. Using the regularity of the
$R$-matrix and eq. \eqref{eq:Gdef} we get
\begin{equation}
 \check \Ga_{123}(v,v) = 1.
\end{equation}
Using the factorization of the $R$-matrix and eqs. \eqref{eq:RG1} and \eqref{eq:RG2} we obtain that
\begin{align}
 \check \Ga_{123}(u,0) &= \check \La_{123}(u),\\
 \check \Ga_{123}(0,v) &= \check \La_{123}(v)^{-1}.
\end{align}
From \eqref{eq:Gdef} we can also derive the inversion property of the $G$-operator
\begin{equation}
 \check \Ga_{123}(u,v) \check \Ga_{123}(v,u) = 1.
\end{equation}
The equations \eqref{eq:RG1},\eqref{eq:RG2},\eqref{eq:Gdef} and \eqref{eq:Gdef2} imply that the $R$- and $G$-operators act diagonally on the first and last sites.

We also know that the $R$-matrix satisfies the YB equation
\begin{multline}
\check{R}_{34,56}(u_{1},u_{2})\check{R}_{12,34}(u_{1},u_{3})\check{R}_{34,56}(u_{2},u_{3})=\\
\check{R}_{12,34}(u_{2},u_{3})\check{R}_{34,56}(u_{1},u_{3})\check{R}_{12,34}(u_{1},u_{2}).
\end{multline}
Substituting \eqref{eq:RG1},\eqref{eq:RG2} we can easily show that the $G$-operator also satisfies a YB type equation
\begin{multline}
\check{\mathcal{G}}_{234}(u_{1},u_{2})\check{\mathcal{G}}_{123}(u_{1},u_{3})\check{\mathcal{G}}_{234}(u_{2},u_{3}) =\\
\check{\mathcal{G}}_{123}(u_{2},u_{3})\check{\mathcal{G}}_{234}(u_{1},u_{3})\check{\mathcal{G}}_{123}(u_{1},u_{2}).
\end{multline}
Below we show that this is equivalent to the ``face weight'' formulation of the Yang-Baxter relation.

To see this we introduce a new notation for the Lax operator, by making use of its special form:
\begin{equation}
  \label{Larep}
\check{\La}(u) =\sum_{i,j,k,l}M_{kj}^{il}(u)P_{i}\otimes E_{\:k}^{l}\otimes P_{j}.
\end{equation}
Assuming that the $G$-operator has a similar form we write it as
\begin{equation}
  \label{Garep}
\check{\Ga}(u,v)	=\sum_{a,b,c,d}(g_{bc})_{d}^{a}(u,v) 
P_{a}\otimes E_{\:b}^{c}\otimes P_{d}.
\end{equation}
Then the GLL relation is expressed as
\begin{multline}
  \label{IRFYB}
\sum_{s}(g_{bs})_{r}^{i}(u,v) M_{sj}^{il}(u)M_{rq}^{sj}(v)= \\
\sum_{s}M_{bs}^{il}(v)M_{rq}^{bs}(u)(g_{sj})_{q}^{l}(u,v),
\end{multline}
which is a relation used in \cite{prosen-cellaut}. This connection will be discussed further in Section \ref{sec:prosen}.

Let us now focus on the quantum circuits. We intend to build brickwork circuits which can accommodate the elementary
cellular automata. It is clear that the constructions discussed in Section \ref{sec:circuit} are not appropriate for
this purpose, because there
the local unitaries are too far away from each other. Instead, we need to build ``tightly packed'' quantum circuits
where every pair of neighboring quantum gates is overlapping at the common 
control bit. This will enable us to treat some of the elementary cellular automata discussed in Section \ref{sec:cellintro}.

Motivated by the discussion in Section \ref{sec:cellintro} we build a Floquet-type time evolution operator  as 
\begin{equation}
  \label{IRFV2}
  \mathcal{V}=\mathcal{V}_2\mathcal{V}_1,
\end{equation}
where the operators $\mathcal{V}_{1,2}$ are built from three site unitaries that we choose as
\begin{equation}
  U^{(3)}(j)=\check \La_{j,j+1,j+2}(\theta)
\end{equation}
The number $\theta$ will be a fixed parameter of the quantum circuit. 

The update steps are then given by
\begin{equation}
  \begin{split}
  \mathcal{V}_1 &= \check{\La}_{1,2,3}(\theta) \check{\La}_{3,4,5}(\theta) \dots 
 \check{\La}_{L-3,L-2,L-1}(\theta)  \check{\La}_{L-1,L,1}(\theta) \\
\mathcal{V}_2 &= \check{\La}_{2,3,4}(\theta) \check{\La}_{4,5,6}(\theta) \dots 
 \check{\La}_{L-2,L-1,L} (\theta) \check{\La}_{L,1,2}(\theta).
  \end{split}
\end{equation}

\begin{figure}
\centering
\includegraphics[width=1.0\columnwidth]{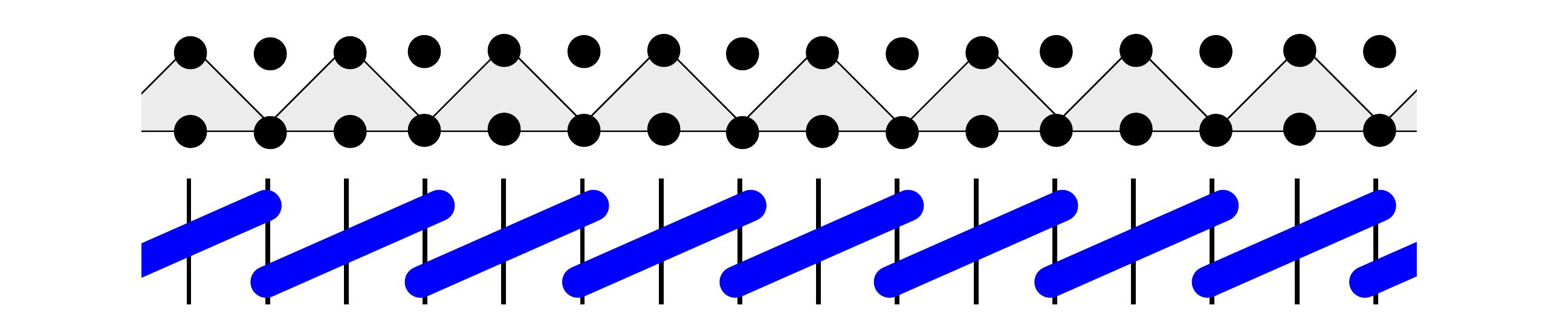}
\caption{Lax representation of the time evaluation operator $\mathcal V_2$.}
\label{fig:LaxrepFo}
\end{figure}

Notice that every pair of neighboring Lax operators overlaps at a common site. For a graphical interpretation see
see Fig. \ref{fig:LaxrepFo}.

\begin{figure}
\centering
\includegraphics[width=1.0\columnwidth]{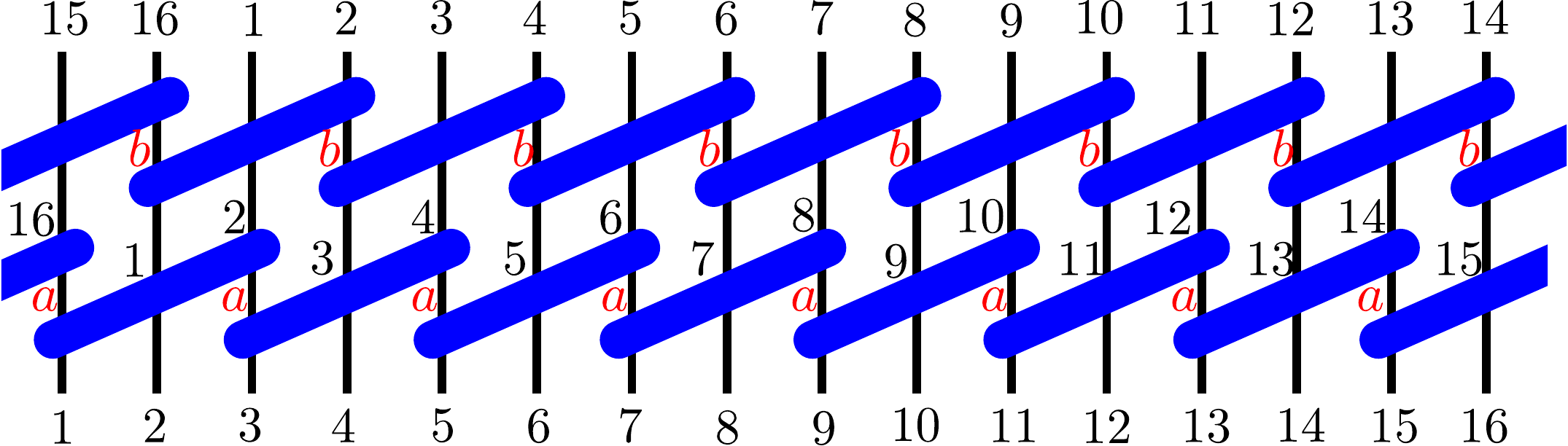}
\caption{Lax representation of the time evaluation operator $\mathcal V_2\mathcal V_1$.}
\label{fig:graphProof}
\end{figure}

The complete Floquet time step can be expressed alternatively as
\begin{multline}
   \mathcal{V}= \mathcal{U}^{2}
 \mathrm{tr}_{ab} \bigl[
 \La_{1,2,b} \La_{1,2,a} \La_{3,4,b} \La_{3,4,a} \dots \\
  \La_{L-3,L-2,b} \La_{L-3,L-2,a} \La_{L-1,L,b} \La_{L-1,L,a}\bigr].
\end{multline}
For simplicity we suppressed the dependence on $\theta$.
For a graphical proof of the rewriting see Fig. \ref{fig:graphProof}.

We can now use the factorization of the $R$-matrix, which we write in the form
\begin{equation}
  \label{tRfact}
 R_{(12),(34)}(\theta,0) = \La_{1,2,4}(\theta)\La_{1,2,3}(\theta).
\end{equation}
Using this relation we can express the Floquet update step as
\begin{multline}
 \mathcal{V} = \mathcal{U}^{2}
 \mathrm{tr}_{A} \bigl[
 R_{(1,2),A}(\theta,0) R_{(3,4),A}(\theta,0) \dots \\
 R_{(L-3,L-2),A}(\theta,0) R_{(L-1,L),A}(\theta,0) \bigr],
\end{multline}
where $A=(a,b)$ stands for a pair of auxiliary spaces. Defining the transfer matrix
\begin{multline}
  \label{eq:IRFtransfer}
 t(u) =
 \mathrm{tr}_{A} \bigl[
 R_{(1,2),A}(\theta,u) R_{(3,4),A}(\theta,u) \dots \\
 R_{(L-3,L-2),A}(\theta,u) R_{(L-1,L),A}(\theta,u) \bigr]
\end{multline}
we obtain that
\begin{equation}
 \mathcal V = t(\theta)^{-1} t(0),
\end{equation}
where we used that
\begin{equation}
 t(\theta) = \mathcal U^{-2}.
\end{equation}
The transfer matrices \eqref{eq:IRFtransfer} form a commuting family:
\begin{equation}
  [t(u),t(v)]=0.
\end{equation}
The variable $\theta$ plays the role of an inhomogeneity parameter for this commuting family. Note that
\eqref{eq:IRFtransfer} is a generalization of \eqref{eq:transferRev}, and not of \eqref{eq:transfer}.

We can also define the rapidity dependent update rule
\begin{equation}
   \mathcal V(u) = t(\theta)^{-1} t(u).
\end{equation}
Clearly, these operators commute with each other, and thus they commute also with the ``physical'' update step which is
obtained at $u=0$.

The operators $\mathcal V(u)$ are non-local for generic $u$. However, similar to the case of a standard transfer matrix
they lead to extensive and local charges. The initial condition $\mathcal{V}(\theta)=1$ and the
definition of the transfer matrix $t(u)$ lead to the extensive four-site operator
\begin{equation}
  \label{Q4v}
  Q'_4=\left.\partial_u   \mathcal V(u)\right|_{u=\theta}.
\end{equation}
The transfer matrix is only two-site invariant, which is inherited by $Q_4'$. Therefore $Q_4'$ is not translationally
invariant, similar to the update rule $\mathcal{V}$ which is not translationally invariant either.

Higher charges could be obtained from the derivative of the logarithm of the transfer matrix, or with the boost operator method
\cite{hubbard-boost} 
applied to transfer matrix \eqref{eq:IRFtransfer}.

\begin{figure}
\centering
\includegraphics[width=1.0\columnwidth]{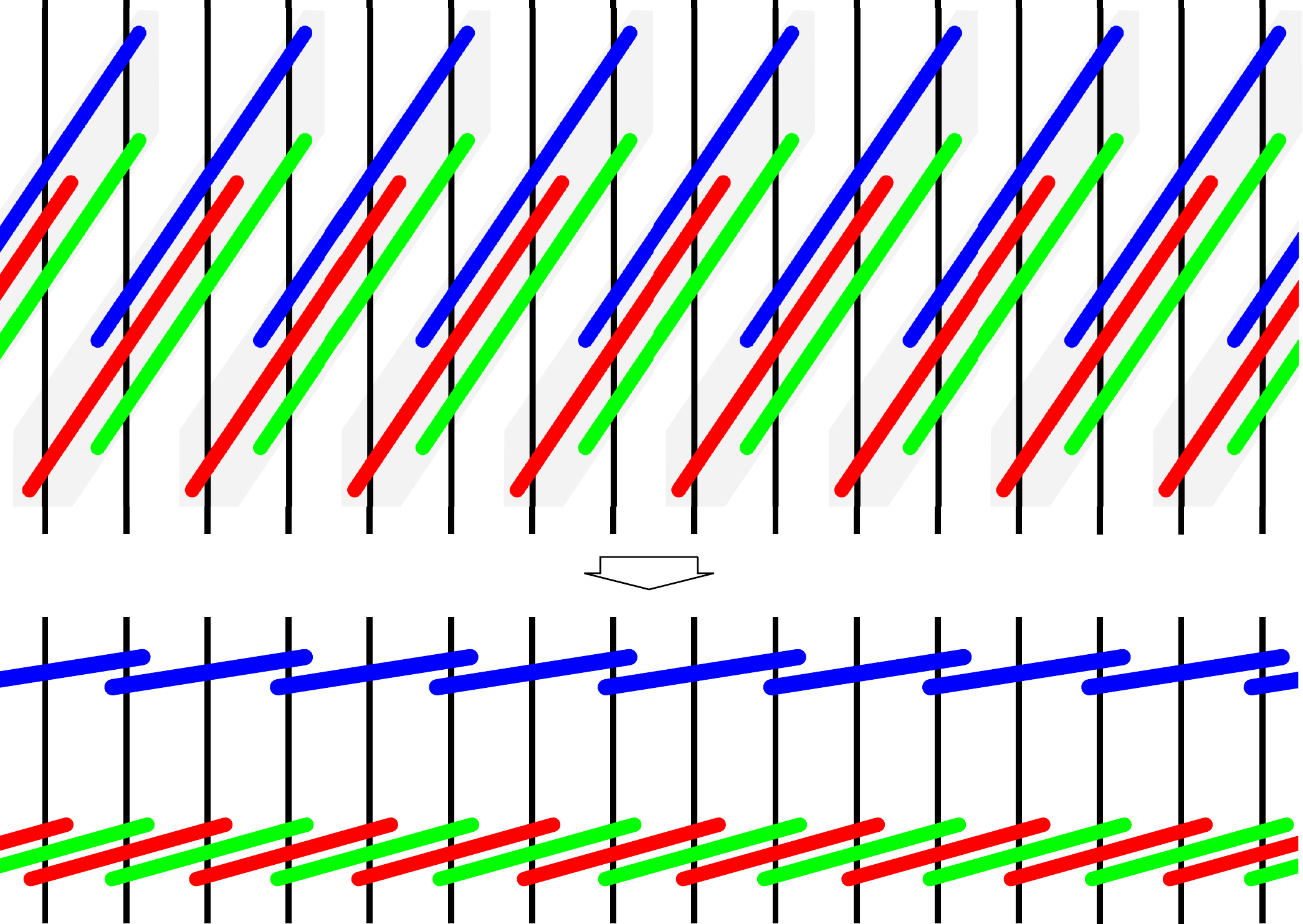}
\caption{Graphical illustration of the transfer matrix \eqref{eq:IRFtransfer}. The blue, red and green boxes are the
  operators $\check \La(\theta)$, $\check \Ga(\theta,u)$ and $\check{\La}(u)^{-1}$, respectively. We use the
  commutativity \eqref{IRFLcomm} to separate the action of the upper (blue) layer of gates. However, the lower layer can
not be simplified further, because there the consecutive operators overlap at two sites each.}
\label{fig:transzWthG}
\end{figure}

Simplified formulas for the transfer matrix can be obtained using the $\check \Ga$ operators introduced above.
We substitute the expression
\begin{equation} \label{eq:RGL}
 \check{R}_{12,34}(\theta,u) = \check{\La}_{234}(\theta)
 \check{\Ga}_{123}(\theta,u) \check{\La}_{234}(u)^{-1}
\end{equation}
into the definition \eqref{eq:IRFtransfer}.
This substitution is depicted pictorially in the top of the figure \ref{fig:transzWthG}.
We can see here that the transfer matrix can be written as 
\begin{equation} 
\label{eq:transferIRF}
t(u) = \mathcal{V}_2 \tilde{t}(u),
\end{equation}
where
\begin{multline}
  \label{sajatI}
 \tilde t(u) = \\
 \mathrm{tr}_{a,b} \bigl[
 \Ga_{1,a,b}(\theta,u) \La_{2,a,b}(u)^{-1}
 \Ga_{3,a,b}(\theta,u) \La_{4,a,b}(u)^{-1} \dots \\
 \Ga_{L-1,a,b}(\theta,u) \La_{L,a,b}(u)^{-1} \bigr].
\end{multline}
For this rewriting we used again the fact that the $\check \La$ operators commute if they share a control bit.

The operator $\tilde t(u)$ is completely identical with the transfer matrices defined in \cite{prosen-cellaut}, which can
be seen after proper identifications are made. To this order we need to use the representations \eqref{Larep} and
\eqref{Garep} for the $\check\La$ and $\check\Ga$ matrices. After substitution we obtain a formula identical to the one
presented in \cite{prosen-cellaut}, see eq. \eqref{prosenI} below.

Our derivation in this Subsection assumed the regularity condition for the $R$-matrix, which leads to the factorization
\eqref{tRfact}. However, an alternative derivation is also possible, without this assumption. We could start with the
definition \eqref{eq:IRFtransfer} for a proper $R$-matrix, and we then could
still derive the alternative form \eqref{eq:transferIRF}. In concrete cases it could be shown that this transfer matrix
is related to cellular automata. Such a derivation would completely bypass the requirement of the regularity. However,
in this work we are interested in cases which yield local conserved charges. Furthermore we found that if the regularity
condition does not hold then the concrete models do not have more conserved charges at all, see the example of the Rule54
model in \ref{sec:prosen}. Therefore we do not discuss solutions without the regularity condition.

\subsection{Partial classification}

\label{sec:IRFH}

Here we perform a partial classification of three site models, where the Lax operators have the special structure
\eqref{IRFLax}. The methods for the classification are essentially the same as in Section \ref{sec:int}. We are looking
for Lax operators satisfying the RLL relations, and we perform a classification based on the Hamiltonian densities that
are derived from the Lax operators: we apply  the
generalized Reshetikhin condition to find the integrable cases.

It is important the Hamiltonians that we find this way
don't commute with the transfer matrices defined in the previous Subsection, because the transfer matrices involve the
inhomogeneities as well.
The Hamiltonians only commute with homogeneous transfer matrices as defined in \eqref{eq:transfer3site}.
However, we can also regard the  Hamiltonians we find below as new integrable models on their own right.

It follows from the Ansatz \eqref{IRFLax} and the derivation rule  \eqref{Lderiv} that
the Hamiltonians in question have the structure given by \eqref{IRFh} with the action matrices being
\begin{equation}
  h^{ab}=\left.\partial_u f^{ab}(u)\right|_{u=0}.
\end{equation}
Therefore we need to classify three-site Hamiltonians with the particular structure given by \eqref{IRFh}, where the
outer two spins act as control 
bits, and the middle spin is an action bit. For this special class of models we assume that the inversion relation
\eqref{Linversion} holds, therefore we exclude the possibility of having a non-zero $\tilde h_{j,j+1,j+2}$ in the construction
of $Q_5$, see \eqref{q5d}.

The Ansatz \eqref{IRFh} has a total number of 16 real parameters, because there are 4 $g$-matrices which are Hermitian.
 Subtracting the identity component and performing a $U(1)$ rotation the number of
parameters could be narrowed down to 14. However, we found this parameter space to be too big for a first classification
attempt. Instead, we selected an
even more restricted Ansatz given explicitly by
\begin{multline}
  \label{IRFAnsatz}
  h_{123}=A \sigma^x_2+B \sigma^z_2+C\sigma^z_1 \sigma^x_2+D \sigma^x_2\sigma^z_3+\\
  +E\sigma^z_1\sigma^z_2\sigma^z_3+
  F\sigma^z_1\sigma^x_2\sigma^z_3+G \sigma^z_1\sigma^z_3.
\end{multline}
Apart from an overall multiplicative normalization this Ansatz has 6 free parameters, which makes the classification of
integrable cases relatively easy. A Hermitian operator is obtained if all parameters are real, and in this case all
matrix elements are real in the computational basis. 

Within this parameter space we found three non-trivial integrable models.
Now we list these models together with a brief discussion of their main properties. Their application as quantum gate
models and classical cellular automata is discussed later in Subsection \ref{elementarycellaut}; here we focus on their
properties as integrable Hamiltonians.

We put forward that two of the models can be related to nearest neighbor chains by a bond-site
transformation. This transformation was used recently in \cite{sajat-folded,sajat-cellaut} and it is explained in detail in
Appendix \ref{sec:bond}. 

\subsubsection{The bond-site transformed XYZ model}

In this case the Hamiltonian density is written in the compact form
\begin{equation}
  \label{eq:BIM}
  h_{123}=
  J_x \sigma^x_2 -  
  J_y \sigma^z_1\sigma^x_2\sigma^z_3+
  J_z \sigma^z_1\sigma^z_3.
\end{equation}
This model can be seen as the bond-site transformed version of the XYZ model (we suggest to call it the bXYZ model). The
Hamiltonian satisfies the requirements 
of the bond-site transformation discussed in Appendix \ref{sec:bond}: it is spin reflection invariant and the first and
last bits are control bits. Using the formulas \eqref{bond-site-op} for the transformation of the operators we see
immediately that in the bond picture the model becomes identical to the XYZ model with the couplings as given
above. Thus it describes interacting dynamics of Domain Walls, where the creation and annihilation of pairs of DW's
is also allowed.

A special point of the model is when $J_x=J_y$, which becomes the bond-site transformation of the XXZ model. In this
case the Hamiltonian commutes with the $U(1)$-charge
\begin{equation}
  \label{Q2zz}
  Q_2=\sum_j \sigma^z_j \sigma^z_{j+1},
\end{equation} 
which is interpreted as the Domain Wall number, which is conserved. This model was already presented in
\cite{pollmann-bond-site-xxz,clifford1}. 

An other special case is when $J_y=0$. In this case the model describes two decoupled quantum Ising chains on the even
and odd sub-lattices, such that the parameter $J_x$ can be interpreted as a magnetic field. Switching on $J_y\ne 0$ we
obtain a Bariev-type coupling between the two sub-lattices, therefore we could also call this system the Bariev-Ising model. 

The Lax operator is found using the known solution for the XYZ model and the bond-site transformation:
\begin{multline}
  \label{ans1}
  \check\La_{1,2,3}(u)= 
  \frac{1}{2}\frac{\sn(\eta)+\sn(u)  \sigma_2^x}{\sn(\eta)+\sn(u)}
  (1-\sigma_1^z\sigma_3^z) + \\
  \frac{1}{2}\frac{\sn(u+\eta)+k\sn(\eta)\sn(u)\sn(u+\eta)\sigma_2^x}{\sn(\eta)+\sn(u)}
  (1+\sigma_1^z\sigma_3^z),
\end{multline}
where $\sn(u)=\sn(u;k)$ and
\begin{equation}
\frac{J_y}{J_x} = \frac{1-k\sn^2(\eta)}{1+k\sn^2(\eta)} \qquad
\frac{J_z}{J_x} = \frac{\cn(\eta)\dn(\eta)}{1+k\sn^2(\eta)}.
\end{equation}

This Lax operator satisfies the inversion relation \eqref{L3inv}
\begin{equation}
  \check\La_{1,2,3}(u)\check\La_{1,2,3}(-u) = 1.
\end{equation}

In this model the $G$-operator can be obtained from the Lax operator in a very natural way
\begin{equation}
 \check \Ga_{123}(u,v) = \check \La_{123}(u-v). 
\end{equation}

Taking $k\to0$ we obtain the XXZ limit of the model
\begin{multline}
  \label{BIlax2}
  \check\La_{1,2,3}(u)=
  (P_1^\circ P_3^\circ+P_1^\bullet P_3^\bullet)
  \frac{\sin(u+\eta)}{ \sin(u)+\sin(\eta)} +\\
+  (P_1^\circ P_3^\bullet+P_1^\bullet P_3^\circ)
\frac{\sin(u) \sigma^x_2+\sin(\eta)}{ \sin(u)+\sin(\eta)}.
\end{multline}
We can also take the XXX limit as $\eta\to 0$ and $u=\eta v$. We obtain the following Lax matrix 
\begin{multline}
  \label{BIlax3}
  \check\La_{1,2,3}(v)=
  (P_1^\circ P_3^\circ+P_1^\bullet P_3^\bullet)+\\
+  (P_1^\circ P_3^\bullet+P_1^\bullet P_3^\circ)
\frac{v \sigma^x_2+1}{v+1}.
\end{multline}

\subsubsection{Twisted XX model with n.n.n. coupling}

\label{sec:twistedXX}

In this model the Hamiltonian density is
\begin{equation}
  \label{2A2B}
  h_{123}=
    \sigma^z_1 \sigma^x_2+\kappa \sigma^x_2\sigma^z_3+G \sigma^z_1\sigma^z_3.
\end{equation}
These are actually two different models depending on the sign $\kappa=\pm 1$; the real parameter $G$ is a coupling constant.
The form of the Hamiltonian density respects the
Ansatz 
\eqref{IRFAnsatz}, but perhaps a more familiar way of writing the global Hamiltonian is
\begin{equation}
  H=\sum_j    \sigma^x_j \sigma^y_{j+1}\pm \sigma^y_j\sigma^x_{j+1}+G \sigma^x_j\sigma^x_{j+2}.
\end{equation}
Here we performed a rotation such that the Pauli matrices are cyclically exchanged as
$\sigma^x\to\sigma^y\to\sigma^z$. In the case of a minus sign we can further express this as
\begin{equation}
  H=\sum_j    2i(\sigma^+_j \sigma^-_{j+1}- \sigma^-_j\sigma^+_{j+1})+G \sigma^x_j\sigma^x_{j+2},
\end{equation}
which can be interpreted as a twisted XX model with a next-to-nearest neighbor interaction term.

In the case of a plus sign in \eqref{2A2B} we do not get such an interpretation.

An alternative interpretation of the model is found by performing a bond-site transformation. First we perform a
transformation $\sigma^x\leftrightarrow\sigma^y$  so that the Hamiltonian density becomes
\begin{equation}
  h_{123}=
    \sigma^z_1 \sigma^y_2+\kappa \sigma^y_2\sigma^z_3+G \sigma^z_1\sigma^z_3.
\end{equation}
This operator is spin flip invariant therefore we can apply the bond-site transformation discussed in Section
\ref{sec:bond}.
Then we obtain a two site interacting model with Hamiltonian density
\begin{equation}
  \label{h12b}
  h_{12}=
    \sigma^y_1 \sigma^x_2+ \kappa \sigma^x_1\sigma^y_2 + G \sigma^z_1\sigma^z_2.
\end{equation}
If $\kappa=-1$, 
then the original Hamiltonian commutes with the $U(1)$-charge given by \eqref{Q2zz}, and the model defined by
\eqref{h12b} commutes with the global $S^z$ operator.
In fact, this case can be interpreted as  the XXZ model with a homogeneous twist, where the kinetic term is the
Dzyaloshinskii–Moriya interaction term. Thus the model describes the interacting dynamics of the Domain Walls, with a
twisted kinetic term.

For the model with $\kappa=1$ we did not find a translationally invariant $U(1)$-charge, and the kinetic term describes
the creation and annihilation of pairs of Domain Walls.

We found the Lax operator for both models. It is given by
\begin{multline}
  \check\La_{123}(u) =
    G \frac{e^{2u}-1}{(e^u-1)^2-4G^2} \times \\
    \left( \frac{2G\kappa}{e^u-1}+
    \left(\sigma^z_1 \sigma^x_2 +\kappa \sigma^x_2\sigma^z_3\right)+
    \frac{e^u-1+2G^2}{G(e^u+1)} \sigma^z_1\sigma^z_3 \right).
\end{multline}
Relation \eqref{L3inv} holds with this parametrization. This operator is unitary if $u$ is purely imaginary. 

The $G$-operator (which immediately gives the $R$-matrix as well)
reads as
\begin{multline}
  \check \Ga_{123}(u,v) =
     \kappa\frac{(4G^2-1)e^v-e^u+e^u e^v+1}{2G(e^u-e^v)}+\\
    \left(\sigma^z_1 \sigma^x_2 +\kappa \sigma^x_2\sigma^z_3\right) +
    G\frac{(4G^2-3)e^v+e^u+e^u e^v+1}{(2G^2-1)(e^u+e^v)+e^u e^v+1} \sigma^z_1\sigma^z_3.
\end{multline}

Substituting the special point $G=1$ we get
\begin{multline}
  \label{model2G1}
  \check\La_{123}(u) =
    \frac{e^{u}-1}{e^u-3} \times \\
    \left( \frac{2\kappa}{e^u-1}+
    \left(\sigma^z_1 \sigma^x_2 +\kappa \sigma^x_2\sigma^z_3\right)+
    \sigma^z_1\sigma^z_3 \right),
\end{multline}
A special point is $u=i\pi$, at which the unitary operator becomes
deterministic; this will lead to an elementary cellular automata, see Section \ref{elementarycellaut} below.

\subsubsection{Integrable deformation of the PXP model}

This model has no free parameters, just a sign $\kappa=\pm 1$:
\begin{equation}
  \label{PXP}
  h_{123}=\sigma^x_2+\kappa(\sigma^z_1 \sigma^x_2+\sigma^x_2\sigma^z_3)+\sqrt{2}\sigma^z_1\sigma^z_2\sigma^z_3
  -\sigma^z_1\sigma^x_2\sigma^z_3.
\end{equation}
In the case of $\kappa=1$  the Hamiltonian density can be expressed as
\begin{equation}
    h_{123}=
  -4P^\bullet_1\sigma^x_2P^\bullet_3+2\sigma^x_2 +\sqrt{2}\sigma^z_1\sigma^z_2\sigma^z_3,
\end{equation}
while for $\kappa=-1$ we would obtain a similar model with $P^\bullet$ replaced by $P^\circ$.
These models can be seen as an integrable deformation of the PXP model \cite{PXP}. However, if we remain in the
parameter space 
of our Ansatz, then there is no free parameter, so we can not tune the ``operator distance'' from the PXP Hamiltonian.

It is likely that the models \eqref{PXP} are just particular cases of a continuous family of models which stretches
outside our Ansatz. We leave the exploration of the bigger parameter space to future works. 

For this model we find the Lax operator
\begin{equation}
 \check \La_{123}(u) = \frac{1 + u h_{123}}{1+\sqrt{6} u}.
\end{equation}
Simple computation shows that 
  \begin{equation}
    (h_{123})^2=6.
\end{equation}
This implies that \eqref{L3inv} is satisfied. In this case a unitary gate is obtained for $u$ being purely imaginary.

For this specific model we did not find a bond-site transformation, which would make it locally equivalent to a n.n. chain.

\subsection{Elementary cellular automata}

\label{elementarycellaut}

The construction of Subsection \ref{sec:IRFgates} can be applied for every IRF type 3-site model
treated in Section  \ref{sec:IRFH}: this way we obtain families of quantum cellular automata with varying numbers of free
parameters. All of these models are Yang-Baxter integrable. Now we are looking for specific cases that can accommodate the
elementary cellular automata discussed in Section \ref{sec:cellintro}. 

First we consider the bXYZ model \eqref{eq:BIM} at the special point $J_x=J_y=J_z$, in which case the Lax operator is given by
\eqref{BIlax3}. Taking the $v\to \infty$ limit we obtain the three site unitary
\begin{multline}
  \label{cellautM1s}
    U^{(3)}(1)
   = (P^\bullet_1P^\circ_3+P^\circ_1P^\bullet_3)\sigma^x_2 +  P^\bullet_1P^\bullet_3+P^\circ_1P^\circ_3.
\end{multline}
We recognize that this is the update rule for the classical Rule150 model given by  \eqref{eq:rule150}. Thus we
obtained a Yang-Baxter integrable three parameter family of quantum cellular automata (with the parameters being
$J_y/J_x$, $J_z/J_x$ and $v$, or alternatively $k$, $\eta$
and $v$), which includes the Rule150 model at special points. At this special point the first non-trivial local conserved charge is
\begin{multline}
  \label{Q4v1}
 Q'_4 =  \sum_{j} \sigma_{2j}^x \sigma_{2j+1}^x +
 \sigma_{2j-1}^z \sigma_{2j}^y \sigma_{2j+1}^y \sigma_{2j+2}^z +\\
+ \sigma_{2j-1}^z \sigma_{2j}^z \sigma_{2j+1}^z \sigma_{2j+2}^z.
\end{multline}

Let us now perform the bond-site transformation of Appendix \ref{sec:bond} directly on the quantum cellular automata. In
the bond picture we obtain two-site gates that are given directly by the $R$-matrices of the XXX, XXZ and XYZ
models. In the XXZ case the two-site gate is given by  \eqref{XXZRcheck}, unitary
time evolution is obtained if $\eta\in\valos$ and $v\in i\valos$ or vice versa. The classical Rule150 model is obtained
after taking the special limits $\eta\to 0$ and $v\to \infty$, in which case the $R$-matrix \eqref{XXZRcheck} becomes a
permutation operator (swap gate). This implies that the Rule150 model describes free movement of Domain Walls. The XXZ
version can be seen as an interacting deformation where the total number of DW's is conserved. Finally the XYZ case is
the most general model in this family, where Domain Walls can be created or annihilated in pairs.

Let us also consider the Model of Section \ref{sec:twistedXX} with the sign $\kappa=1$ and the coupling constant $G=1$,
for which the Lax operator and thus the three-site unitary is given by \eqref{model2G1}. 
Further substituting $u=i\pi$ we obtain
\begin{multline}
  \label{model2G1b}
  U^{(3)}(1) =
     \frac{1}{2} 
    \left( -1+
\sigma^z_1 \sigma^x_2 +\sigma^x_2\sigma^z_3+
    \sigma^z_1\sigma^z_3 \right),
\end{multline}
This is also a deterministic quantum gate, which gives the $f$ matrices
\begin{equation}
   \label{eq:rule105b}
  f^{00}=-f^{11}=\sigma^x, \quad f^{01}=f^{10}=-1.
\end{equation}
We can see that apart from simple signs these $f$-matrices are equal to those of the Rule105 model given by 
\eqref{eq:rule105}. As already argued in \cite{sajat-cellaut}, if the quantum gates are deterministic, then the phases are
irrelevant for the simulation of a classical cellular automata. Thus the two-parameter family of quantum gates
\eqref{model2G1} can be considered as a deformation of the actual Rule105 model, which is included at the special point
$G=1$ and $u=i\pi$.

For this model and the specific values $G=1$ and $u=i\pi$ the definition \eqref{Q4v} gives the local conserved charge
\begin{multline}
  \label{Q4v2}
 Q'_4 = 
 \sum_{j} \sigma_{2j}^y \sigma_{2j+1}^y 
+ \sigma_{2j-1}^z \sigma_{2j}^x \sigma_{2j+1}^x \sigma_{2j+2}^z +\\
+ \sigma_{2j-1}^z \sigma_{2j}^z \sigma_{2j+1}^z \sigma_{2j+2}^z.
\end{multline}
Performing the transformation $\sigma^x\leftrightarrow\sigma^y$ mentioned above we obtain the same charge as in
\eqref{Q4v1}. 

The charges \eqref{Q4v1} and \eqref{Q4v2} commute with the time evolution of the update rules given by
\eqref{eq:rule150} and \eqref{eq:rule105b}, respectively. The update rules are
deterministic, therefore if we choose a ``classical'' initial state (an element of the computational basis) then it will
stay classical. This also means that for the classical time evolution we can disregard the 
the $\sigma^x$ and $\sigma^y$ operators, because their mean values in the classical states are zero.  This leads to the
following classical charge for the Rule105 and Rule150 models: 
\begin{equation}
   Q^{(cl)}_4 = 
   \sum_{j}
    \sigma_{2j-1}^z \sigma_{2j}^z \sigma_{2j+1}^z \sigma_{2j+2}^z.
\end{equation}
This charge is conserved by both classical cellular automata. Taking further derivatives of the logarithm of the
transfer matrix we could obtain further quantum and classical charges for these models. If we consider the quantum
models then the relative signs in \eqref{eq:rule105b} have to be taken into account. So far we have not yet found a
quantum model which would describe the Rule105 model without these signs, but this could be just a limitation of our
Ansatz \eqref{IRFAnsatz}.

Regarding the Rule54 and Rule201 models we did not find any three-site Hamiltonian, which would lead to Lax operators
and transfer matrices
that would actually accommodate these classical cellular automata. This is in accordance with the findings of
\cite{prosen-cellaut}. Regarding the Rule54 model it was claimed in \cite{prosen-cellaut} that there are no
translationally invariant local 
charges up to interaction range $\ell=5$. This clearly shows that for the Rule54 model we can not have a Lax operator
acting on three sites, because it would give a translationally invariant charge $Q_4'$ with range $\ell=4$. The alerted
reader might object that our $Q_4'$ derived in \eqref{Q4v} is invariant with respect to a two site shift
only.
However, in the case of these cellular automata we also have the inversion relations \eqref{Vinv} which implies that
the 
two-site invariant charge $Q_4'$ commuting with $\mathcal{V}_2\mathcal{V}_1$ has to commute with
$\mathcal{V}_1\mathcal{V}_2$ as well, leading eventually to a translationally invariant version of $Q_4'$ 
commuting with 
both  $\mathcal{V}_2\mathcal{V}_1$ and  $\mathcal{V}_1\mathcal{V}_2$. Altogether we reach the Conclusion that the Rule54
model is not Yang-Baxter integrable with three site interactions.

In Subsection \ref{sec:prosen} below we also analyze the recent construction of
\cite{prosen-cellaut} for these 
models, and we reach the same conclusion: the transfer matrices built in \cite{prosen-cellaut}  only have a diagonal
dressing using a known charge and thus do not lead to extra conserved charges.

However, it is known that there is a local conserved charge for the Rule54 model with interaction range $\ell=6$: it is
the Hamiltonian derived in \cite{vasseur-rule54}. This suggests that perhaps the true algebraic background for the model
lies within the family of 6-site interacting models. We return to this question in the Discussions (Section \ref{sec:disc}).

\subsection{Discussion of the results of \cite{prosen-cellaut} for the IRF models}

\label{sec:prosen}

Recently an algebraic framework for the integrability of the Rule54 and related models was proposed in
\cite{prosen-cellaut}. Here we review this construction, we point out connections with our results, and we also disprove
some of the conjectures made in \cite{prosen-cellaut}.

Let us start with a brief discussion of integrable 2D statistical physical models. 
There are three different sorts of commonly used models: spin models, vertex models, and interaction round a face (IRF)
models. Accordingly, there are three different types of Yang-Baxter equations, one for each family. The different
formulations can always be transformed into each other, although this might not be convenient and it can lead to an
increase in the local dimensions. For a summary of the different formulations see the introductory Sections of
\cite{star-triangle-review1,star-triangle-review2}.

Nowadays the most commonly used formulation is the one based on vertex models; this is also what was used throughout the
present work. It was a new idea of  \cite{prosen-cellaut} to construct transfer matrices for the classical cellular
automata (and for certain deformations thereof) using the IRF language.

Now we review the construction of  \cite{prosen-cellaut}, by focusing on the Rule54 model without deformation. We will
show that the transfer matrices of  \cite{prosen-cellaut} have an identical structure as in our quantum circuits discussed above.

The time evolution operator of  \cite{prosen-cellaut} is built exactly in the same way as in \eqref{IRFV}-\eqref{IRFVj}
with the three site gates having the structure given by \eqref{IRFU}. Afterwards a commuting family of transfer matrices
is built using two matrices $L^{j_1,j_2}_{i_1,i_2}(\lambda)$ and $M^{j_1,j_2}_{i_1,i_2}(\lambda)$ which describe {\it
  face weights} in the IRF language. Both matrices have four indices ranging from 1 to 2, and they can be represented
most easily as 
$4\times 4$ matrices using the conventions for the tensor product:
\begin{equation}
L=\left(\begin{array}{cccc}
L_{0,0}^{0,0} & L_{0,0}^{0,1} & L_{0,0}^{1,0} & L_{0,0}^{1,1}\\
L_{0,1}^{0,0} & L_{0,1}^{0,1} & L_{0,1}^{1,0} & L_{0,1}^{1,1}\\
L_{1,0}^{0,0} & L_{1,0}^{0,1} & L_{1,0}^{1,0} & L_{1,0}^{1,1}\\
L_{1,1}^{0,0} & L_{1,1}^{0,1} & L_{1,1}^{1,0} & L_{1,1}^{1,1}
\end{array}\right).
\end{equation}
The number $\lambda$ is interpreted again as a
spectral parameter. The explicit form of the $L$ and $M$ matrices is given by
\begin{equation}
  L(\lambda)=
  \begin{pmatrix}
    1 & & 1 & \\
    1 & & 1 & \\
    & \lambda^2 & & \lambda \\
    & \lambda & & 1 \\    
  \end{pmatrix},
\end{equation}
\begin{equation}
  M(\lambda)=
  \begin{pmatrix}
    1 & & 1 & \\
     & 1/\lambda &  & 1/\lambda \\
    & 1 & 1/\lambda & \\
 1   &  &1/\lambda &  \\    
  \end{pmatrix}.
\end{equation}
The transfer matrices $\mathcal{I}(\lambda)$ are defined component-wise as
\begin{equation}
  \label{prosenI}
  \mathcal{I}^{j_1,j_2,\dots,L}_{i_1,i_2,\dots,i_L}(\lambda)=
  \prod_{x=1}^{L/2} M^{j_{2x-1},j_{2x}}_{i_{2x-1},i_{2x}}(\lambda)
  L^{j_{2x},j_{2x+1}}_{i_{2x},i_{2x+1}}(\lambda).
\end{equation}
There is no summation over repeated indices.

It is then proven in \cite{prosen-cellaut} that the transfer matrices commute with each other and also with the time
evolution operator:
\begin{equation}
[  \mathcal{I}(\lambda),\mathcal{V}]=0.
\end{equation}
This is shown using the ``face weight'' formulation of the Yang-Baxter relation discussed above.

This  construction of  \cite{prosen-cellaut} is completely identical with our quantum circuits of Section \ref{sec:IRFgates}
after proper identifications are made. It was already shown in Section \ref{sec:IRFgates} that our GLL and GGG relations
are identical to the ``face weight'' formulation of the Yang-Baxter relation given by eq. \eqref{IRFYB}. Furthermore, we recognize
the structural similarity between the transfer matrices \eqref{sajatI} and \eqref{prosenI}. This leads to the following
correspondences: 
\begin{align}
  \check{\mathcal{L}}_{123}(\theta) &\equiv f \\
  \check{\mathcal{L}}_{123}(u) &\equiv M \\
  \check{\mathcal{G}}_{234}(\theta,u) &\equiv L \\
  \check{\mathcal{G}}_{123}(u_{1},u_{2}) &\equiv g,
\end{align}
where on the l.h.s. we listed our operators, and the r.h.s. contains the objects defined in \cite{prosen-cellaut}. For
the identification of the indices see eqs. \eqref{Larep} and \eqref{Garep}.

Let us now show that in the particular case of the Rule54 model the transfer matrix \eqref{prosenI} does not yield new
conserved charges.

First of all we consider two known conserved operators in this model: the translation operator $\UU$ and
the particle current defined as
\begin{equation}
  \label{prosenJ}
  \mathcal{J}=\sum_{x=1}^{L/2} \big(\sigma^z_{2x-1}\sigma^z_{2x}-\sigma^z_{2x}\sigma^z_{2x+1}\big).
\end{equation}
This operator anti-commutes with both the shift and the single step update operators:
\begin{equation}
  \label{Janticomm}
  \{ \mathcal{J},\UU\}=  \{ \mathcal{J},\mathcal{V}_1\}=  \{ \mathcal{J},\mathcal{V}_2\}=0
\end{equation}
It follows that it commutes with the Floquet-cycle operator $\mathcal{V}=\mathcal{V}_2\mathcal{V}_1$:
\begin{equation}
 [ \mathcal{J},\mathcal{V}]=0.
\end{equation}
The translation operator intertwines the two update steps:
\begin{equation}
  \mathcal{V}_1\UU=\UU \mathcal{V}_2,
\end{equation}
Strictly speaking $\UU$ is not conserved by the Floquet cycle $\mathcal{V}$, because it interchanges the odd and even
sites. On the other hand, $\UU^2$ is conserved.

It was conjectured  in \cite{prosen-cellaut} that new quasi-local charges can be obtained from the transfer
matrix \eqref{prosenI}.
On the contrary, we show here that $\mathcal{I}(\lambda)$ is functionally dependent on the time evolution operator and
the two conserved operators  $\UU$ and $\mathcal{J}$.

First we note that the matrix $L^{j_1,j_2}_{i_1,i_2}(\lambda)$ is  diagonal in the indices $j_2\leftrightarrow i_1$, and
it can be factorized as
\begin{equation}
    L^{j_1,j_2}_{i_1,i_2}(\lambda)=\delta_{i_1,j_2}      \lambda^{A^{j_1}_{i_1}+B^{j_2}_{i_2}},
\end{equation}
where
\begin{equation} \label{eq:defAB}
  A^{j_1}_{i_1}=\delta_{j_1,0}\delta_{i_1,1},\qquad
  B^{j_2}_{i_2}=\delta_{j_2,1}\delta_{i_2,0}.
\end{equation}

The factorization above means that the $\lambda$-dependence can be separated into index pairs to the left and to the right. It follows that
the transfer matrix \eqref{prosenI} can be written as
\begin{equation}
  \label{prosenI2}
  \mathcal{I}^{j_1,j_2,\dots,j_L}_{i_1,i_2,\dots,i_L}(\lambda)=
  \prod_{x=1}^{L/2} \tilde M^{j_{2x-1},j_{2x}}_{i_{2x-1},i_{2x}}(\lambda)
  \delta_{i_{2x},j_{2x+1}},
\end{equation}
where
\begin{equation}
 \tilde M^{j_1,j_2}_{i_1,i_2}(\lambda)=
\lambda^{B^{j_1}_{i_1}}  M^{j_1,j_2}_{i_1,i_2}(\lambda)\lambda^{A^{j_2}_{i_2}}.
\end{equation}
Once again there is no summation over repeated indices.

Furthermore, the same operator can be written as $\mathcal{I}(\lambda)=\UU \tilde{\mathcal{I}}(\lambda)$, where now the
components of $\tilde{\mathcal{I}}(\lambda)$ are
\begin{equation}
  \label{prosenI3}
\tilde{\mathcal{I}}^{j_1,j_2,\dots,L}_{i_1,i_2,\dots,i_L}(\lambda)=
  \prod_{x=1}^{L/2} \tilde M^{j_{2x-2},j_{2x-1}}_{i_{2x-1},i_{2x}}(\lambda)
   \delta_{i_{2x},j_{2x}}.
\end{equation}
We can see that this operator acts as the identity on the even sites, which become control bits for the action on the
odd sites. To be more precise, the same operator can be written as a product of commuting three-site unitaries
\begin{equation}
  \tilde{\mathcal{I}}(\lambda)=\prod_{x=1}^{L/2}  U^{(3)}(2x|\lambda),
\end{equation}
where $U^{(3)}(2x|\lambda)$ has the form of \eqref{IRFU} with the $\lambda$-dependent $f$-matrices given through the
matrix elements
\begin{equation}
  \big(f^{ab}(\lambda)\big)_{i}^j =\tilde M^{aj}_{ib}(\lambda).
\end{equation}
Considering the concrete components of $\tilde M$ we can write the individual $f$-matrices as
\begin{equation}
  f^{11}=\sigma^x,\quad f^{00}=1
\end{equation}
and
\begin{equation}
   f^{10}=
   \begin{pmatrix}
     \lambda & \\ & 1/\lambda 
   \end{pmatrix}\sigma^x,\quad  f^{01}= \begin{pmatrix}
     1/\lambda & \\ & \lambda 
   \end{pmatrix}\sigma^x.
\end{equation}
Comparing to \eqref{eq:rule54} we see that these are $\lambda$-deformed versions of the original $f$-matrices of the
model. Thus we can write
\begin{equation}
   \tilde{\mathcal{I}}(1)=\mathcal{V}_2,
\end{equation}
where $\mathcal{V}_{1,2}$ are the two operators that define the update rules, see \eqref{IRFV}-\eqref{IRFVj}.

Collecting the factors of $\lambda$ as we multiply the equal time quantum gates we obtain
\begin{equation}
  \tilde{\mathcal{I}}(\lambda)=\lambda^{\mathcal{J}/2}\ \mathcal{V}_2,
\end{equation}
where $\mathcal{J}$ is the particle current operator defined in \eqref{prosenJ}.
Going back to the actual transfer matrix we get
\begin{equation}
  \mathcal{I}(\lambda)=\UU\ \lambda^{\mathcal{J}/2}\ \mathcal{V}_2.
\end{equation}
This formula means that the transfer matrix $\mathcal{I}(\lambda)$ is functionally dependent on three known operators:
the cyclic shift, the conserved particle current, and the single step update rule. From this formula it follows that
$\mathcal{I}(\lambda)$ is unitary if $|\lambda|=1$; this was an unexplained observation of
\cite{prosen-cellaut}. Finally, for the product of two transfer matrices we obtain
\begin{equation}
 \mathcal{I}(\lambda_2) \mathcal{I}(\lambda_1)= \UU^2 (\lambda_1\lambda_2)^{-\mathcal{J}} \mathcal{V}.
\end{equation}
Here we used the anti-commutation relations \eqref{Janticomm} and the definition of the Floquet cycle
$\mathcal{V}$. This proves the commutativity of the transfer matrices. Furthermore, choosing $\lambda_1=\lambda_2$ we find
that the squared transfer matrix is simply just a combination of the two-site translation, the conserved
particle current, and the Floquet update rule.

We interpret this result as follows: Even though the Rule54 model seems integrable, the construction of
\cite{prosen-cellaut} can not be considered as a proof of it, because \cite{prosen-cellaut} fails to introduce new
charges on top of the existing ones.  
The same conclusion can be reached for the deformed Rule54 model, where the $M$-matrices are modified but the
$L$-matrices are kept the same \cite{prosen-cellaut}, so that the key steps of our computation here can be applied in
the same way.

\section{Four site interactions}

\label{sec:four}

It is relatively straightforward to generalize the results of the previous Sections to models with four site
interactions. In Section \ref{sec:threesite} the key ideas were obtained after we constructed a nearest neighbor chain
by gluing pair of sites together. In the case of four site interactions we need to group together triplets of spins,
thus obtaining a nearest neighbor chain for the glued sites. This n.n. chain is integrable, therefore we expect
that it has a regular $R$-matrix. We have glued together three sites, therefore the auxiliary space for this $R$-matrix
has to be a tensor product of three auxiliary spaces $a,b,c$. We can then construct transfer matrices in an analogous
way as in \eqref{eq:transfer3site} but now with $R_{(a,b,c),(j,j+1,j+2)}(u,0)$ which acts on the triplets of
physical sites and auxiliary spaces.

Going further, we  need to satisfy the condition that the charges of the original chain are translationally invariant,
and as an effect we expect that the transfer matrix will also be translationally invariant. This condition leads to the
factorization of the $R$-matrix as
\begin{multline}
    \label{R4fact}
    R_{(a,b,c),(j,j+1,j+2)}(u,0) =\\
    =\La_{a,b,c,j+2}(u) \La_{a,b,c,j+1}(u) \La_{a,b,c,j}(u).
\end{multline}
Here $\La_{a,b,c,j}(u)$
is the Lax operator acting on three auxiliary spaces and a single physical space.

The transfer matrix is then constructed as
\begin{equation}
  \label{tL4}
  t(u)=\text{Tr}_{a,b,c}\La_{a,b,c,L}(u)\dots \La_{a,b,c,1}(u).
\end{equation}
The regularity condition for the $R$-matrix implies the initial condition
\begin{equation}
\label{Linit2}
  \La_{a,b,c,j}(0)=\Pe_{a,j}\Pe_{b,j}\Pe_{c,j},
\end{equation}
which leads to
\begin{equation}
  t(0)=\UU^3.
\end{equation}
Writing the Lax operator as
\begin{equation}
   \La_{a,b,c,j}(u)=\Pe_{a,j}\Pe_{b,j}\Pe_{c,j} \check  \La_{a,b,c,j}(u)
\end{equation}
we compute the four site Hamiltonian density as
\begin{equation}
   h_{1,2,3,4}=\left. \partial_u \check \La_{1,2,3,4}(u)\right|_{u=0}.
\end{equation}
For the Lax operator we expect the inversion relation
\begin{equation}
  \label{L4inv}
   \check  \La_{a,b,c,j}(u) \check  \La_{a,b,c,j}(-u)=1.
\end{equation}
From this we can compute the next conserved charge from the transfer matrix. It will be a 7-site operator
\begin{equation}
  Q_7=\sum_j   q_7(j)
\end{equation}
with
\begin{equation}
  \label{q7}
  q_7(1)=\left[h_{1,2,3,4},
  \sum_{k=1}^3 h_{1+k,2+k,3+k,4+k}\right].
\end{equation}
The commutativity of $H$ and $Q_7$ can be used as an integrability criterion, which can serve as a starting point for
classifying four site interacting models. 

As an initial step in this direction we classified all $SU(2)$ invariant models with space reflection symmetry. Sorting
out the trivial cases we found only one new model, with the Hamiltonian density being
\begin{equation}
  \label{foursitemodel}
h_{1,2,3,4}=2\left(\Pe_{1,4}-1\right)\left(\Pe_{2,3}-1\right)-\Pe_{1,3}-\Pe_{2,4}.
\end{equation}
Given the huge literature of integrable models we can not be entirely certain that the model has not yet appeared in the
literature, possibly in some other form. In any case it appears to be new.

Going further in the classification, an obvious next step is to consider the $U(1)$-invariant models. This opens up a
bigger parameter space, and we leave its exploration to future works. We note that the folded XXZ model treated
in \cite{folded1,folded2,sajat-folded} belongs to this class, and its Hamiltonian is the four-site charge $Q_4$ of
\eqref{foldedQ}. In the next Subsection we derive an integrable quantum circuit for this particular model.

Finally we stress that (in parallel with the three site interacting case) we were not able to prove the factorization
\eqref{R4fact}, therefore we regard it as a 
conjecture. Furthermore, it is not clear whether all integrable solutions can be put in a form which satisfies the
inversion relation \eqref{L4inv}. We leave these problems to future research.

\subsection{Integrable quantum circuit for the folded XXZ model}

\label{sec:folded}

\begin{figure}
\centering
\includegraphics[width=0.99\columnwidth]{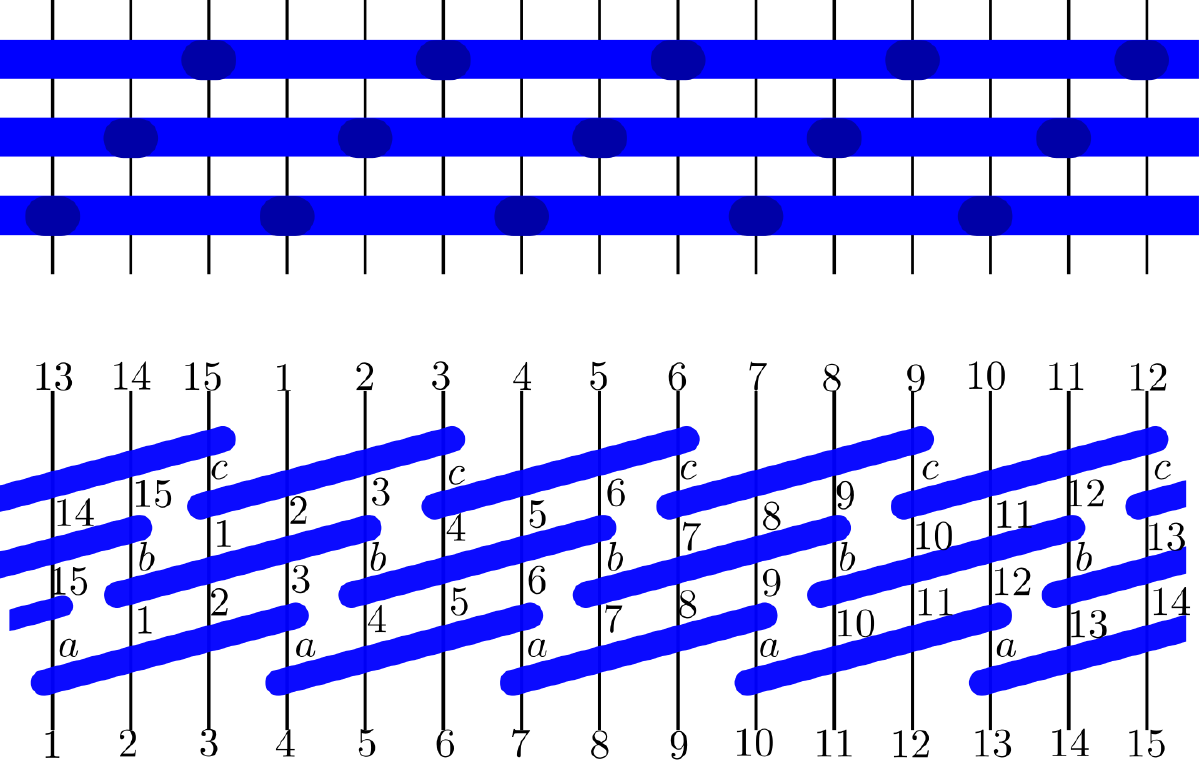}
\caption{Time step operator and its transfer matrix representation. This construction applies to models with four site
  interactions where 
  the outer two spins are control bits. These control bits are depicted as shaded circles in the figure above.
  An example is the folded XXZ model.}
\label{fig:4Qgates}
\end{figure}

A brickwork type quantum circuit for the folded XXZ model was introduced in \cite{sajat-cellaut}. The idea is to build a
Floquet cycle of length $\tau=3$, with four site unitaries $U^{(4)}$ placed at coordinates $x_k=3k$ and with the
displacements $\Delta_l=l$ (see Section \ref{sec:qgatesintro} for the explanation notations). The four-site unitaries
are given by the Lax operator, which reads \cite{sajat-cellaut}
\begin{multline}
  \label{foldedU4}
  U^{(4)}(j|u)=\check \La_{1,2,3,4}(u)=P^\bullet_j P^\circ_{j+3}+P^\circ_j P^\bullet_{j+3} +\\
+\left(P^\bullet_j P^\bullet_{j+3}+P^\circ_j P^\circ_{j+3}  \right) U^{(2)}_{j+1,j+2}(u),
   \end{multline}
   where $U^{(2)}_{j+1,j+2}(u)$ is a two site unitary given by the explicit matrix representation
\begin{equation}
  \label{U2}
  U^{(2)}(u)=
  \begin{pmatrix}
    1 & 0 & 0 & 0 \\
      0 & \sech(u) & i\tanh(u) & 0\\
   0 & i\tanh(u)  &  \sech(u)  & 0\\
    0 & 0 & 0 & 1\\
  \end{pmatrix}.
\end{equation}   
This matrix is obtained simply from the known $R$-matrix of the XX model. Note that \eqref{foldedU4} has the same
structure as the corresponding charge $Q_4$: it has two control bits  and two action bits. As an effect, the unitaries
commute even if they overlap at the control bits. This enables us to build a Floquet cycle which has periodicity 3 both
in the temporal and the spatial directions. 
With this we have completely specified the quantum circuit. For a graphical interpretation see the upper graph in
Fig. \ref{fig:4Qgates}.

In \cite{sajat-cellaut} the integrability of this circuit was established in the bond picture (after performing the
bond-site transformation discussed in \ref{sec:bond}), where the  building blocks are three site unitaries. In
\cite{sajat-cellaut} diagonal-to-diagonal transfer matrices were constructed, in the same way as in Section
\ref{sec:threegates}.

Now we show that there exists a commuting family of row-to-row transfer matrices in the original picture of this model.
This complements the results of \cite{sajat-cellaut}.

First we start with the discussion of the integrability properties of the special class of four site models, where
the Lax operators satisfies an additional condition
\begin{equation}
 [\check \La_{1234}(u),\check \La_{4567}(v)] = 0,
\end{equation}
i.e. the first and the last sites are control bits. The consequence of this property is that there exists an five site
operator $\check \Ga$ for which the $R$-matrix factorizes as 
\begin{align} \label{eq:4siteRG}
 \check R_{123456}(u,v) =& 
  \check \La_{1234}(v)^{-1} \check \Ga_{23456}(u,v) \check \La_{1234}(u)= \nonumber\\
 =& \check \La_{3456}(u) \check \Ga_{12345}(u,v) \check \La_{3456}(v)^{-1}.
\end{align}
The consistency of these factorizations requires the GLL relation
\begin{multline}
 \check \Ga_{23456}(u,v) \check \La_{1234}(u) \check \La_{3456}(v) = \\
 \check \La_{1234}(v) \check \La_{3456}(u) \check \Ga_{12345}(u,v).
\end{multline}

Let us now construct the single step update rule as
\begin{equation}
  \mathcal{V}_1 = \check{\La}_{1,2,3,4}(\theta) \check{\La}_{4,5,6,7}(\theta) \dots 
 \check{\La}_{L-2,L-1,L,1}(\theta),
\end{equation}
where $\theta$ will be a fixed parameter of the quantum circuit. For the Floquet cycle we obtain
\begin{multline}
\mathcal{V} =\mathcal{V}_3\mathcal{V}_2\mathcal{V}_1 = \mathcal{U}^{3}\times \\
 \mathrm{tr}_{abc} \bigl[
 \La_{1,2,3,c}(\theta) \La_{1,2,3,b}(\theta) \La_{1,2,3,a}(\theta) \\
 \La_{4,5,6,c}(\theta) \La_{4,5,6,b}(\theta) \La_{4,5,6,a}(\theta) \dots \\
  \La_{L-2,L-1,L,c}(\theta) \La_{L-2,L-1,L,b}(\theta) \La_{L-2,L-1,L,a}(\theta) \bigr].   
\end{multline}
Applying the factorization formula
\begin{equation}
 \check R_{123456}(\theta,0)
    =\check \La_{3456}(\theta) \check \La_{2345}(\theta) \check \La_{1234}(\theta)
\end{equation}
we can define a transfer matrix (with auxiliary space $A=(a,b,c)$)
\begin{multline}
  \label{eq:IRFtransfer4}
 t(u) =
 \mathrm{tr}_{A} \bigl[
 R_{(1,2,3),A}(\theta,u) R_{(4,5,6),A}(\theta,u) \dots \\
 R_{(L-5,L-4,L-3),A}(\theta,u) R_{(L-2,L-1,L),A}(\theta,u) \bigr],
\end{multline}
which generates the time step as
\begin{equation}
 \mathcal{V} = t^{-1}(\theta)t(0).
\end{equation}
These transfer matrices commute:
\begin{equation}
  [t(u),t(v)]=0.
\end{equation}
With this we have established a commuting family of transfer matrices that includes the update rule of the quantum
circuit at the special point $u=0$.

\begin{figure}
\centering
\includegraphics[width=0.99\columnwidth]{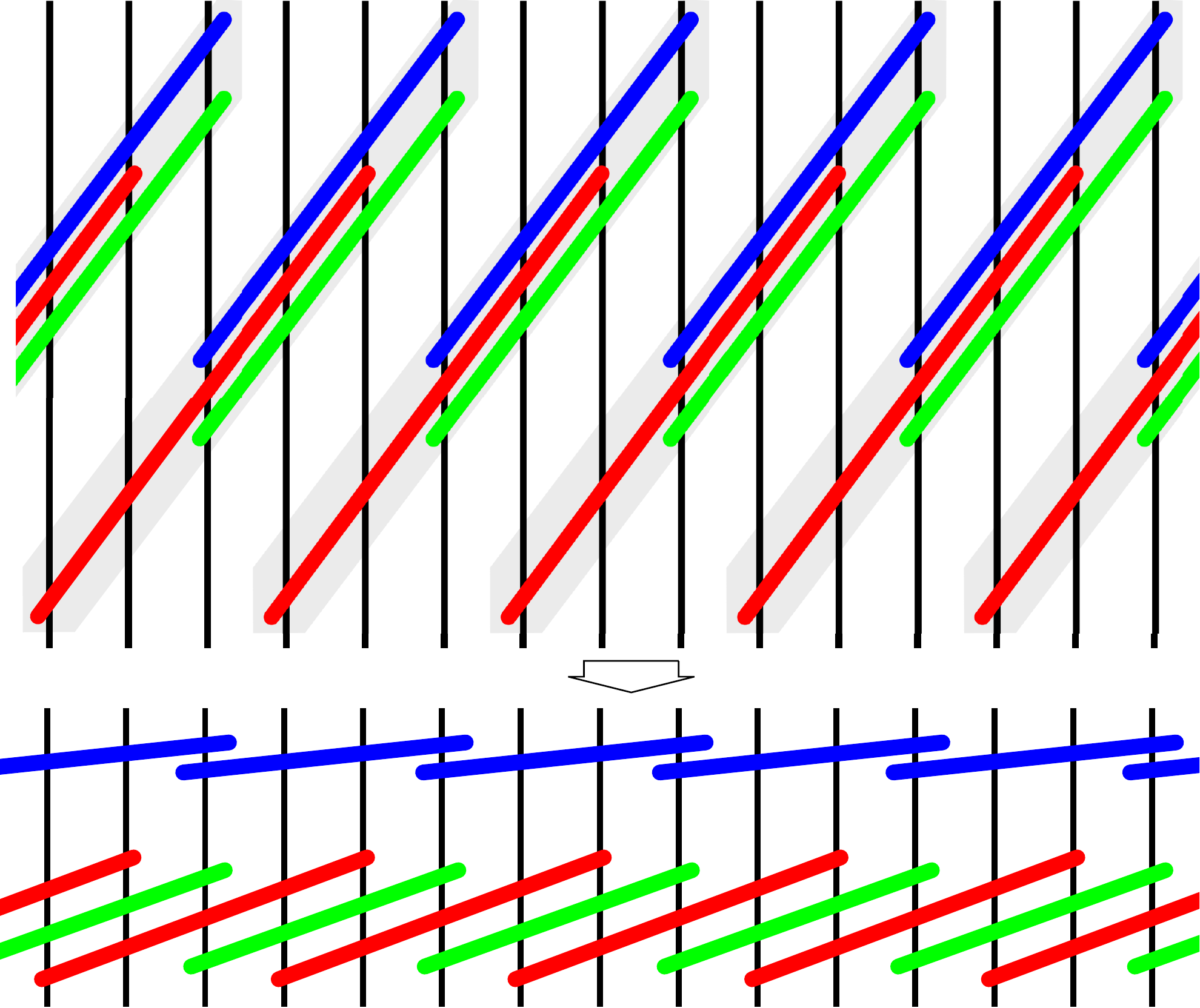}
\caption{Transfer matrix with the operator $\check \Ga$.}
\label{fig:4QgateswG}
\end{figure}

The transfer matrix can be rewritten using the factorization \eqref{eq:4siteRG}. We find
\begin{equation} 
\label{eq:transferIRF4s}
t(u) = \mathcal{V}_3 \tilde{t}(u),
\end{equation}
where
\begin{multline}
  \label{sajatI4}
 \tilde t(u) = \\
 \mathrm{tr}_{a,b,c} \bigl[
 \Ga_{1,2,a,b,c}(\theta,u) \La_{3,a,b,c}(-u)
 \Ga_{4,5,a,b,c}(\theta,u) \La_{6,a,b}(-u)  \\
\dots \Ga_{L-2,L-1,a,b,c}(\theta,u) \La_{L,a,b,c}(-u) \bigr],
\end{multline}
where
\begin{equation}
\Ga_{1,2,3,4,5}(\theta,u) = \Pe_{1,5}\Pe_{2,5}\Pe_{3,5} \Pe_{1,4}\Pe_{2,4}\Pe_{3,4} 
\check \Ga_{1,2,3,4,5}(\theta,u).
\end{equation}

\section{Discussion}

\label{sec:disc}

In this paper we treated integrable spin chains with medium range interaction, focusing on cases with three-site and
four site
interactions. We presented a new algebraic framework which can lead to a classification of such models, and to the
construction of new quantum and classical cellular automata.
As it was explained in the Introduction, one of the most
general problems in the field of integrability is the 
{\it classification of all integrable models}, and clarifying the essential features of integrability. Our results can
be seen as a contribution to this multi decade endeavor. In the paper we treated the three site and four site
interacting models in 
detail, but the generalization to longer interaction ranges is rather straightforward.

We presented partial classifications for  the three site and four site spin-1/2 models, and we found a
number of new models. Given the enormous literature of integrable models it is always difficult to know whether a model
is indeed new; we did our best in the search of the literature and our models appear to be new. We recall that our
models are translationally invariant, and in the three site interacting case they can be pictured as zig-zag spin
ladders. To our best knowledge the only translationally invariant integrable three site chain the literature is the
Bariev model \cite{bariev-model}; other constructions naturally involve a staggering of some of the parameters (see for example
\cite{zigzag1a,zigzag2,zigzag3,zigzag4}) and thus
they are not in the category of models that we are investigating.

In the family of $SU(2)$-invariant spin chains with reflection symmetry we did not find a non-trivial three site model, and we found only one new four site
model given by \eqref{foursitemodel}. In the family of $U(1)$-invariant three site chains (again with space reflection
symmetry) we found two families: the
Bariev model and the hard rod deformed XXZ model given by \eqref{Hhrdef2}. The latter will be analyzed in detail in an
upcoming publication.

A further interesting family of models is that of the IRF
type Hamiltonians and quantum gates. These theories are very similar to known Restricted Solid on Solid (RSOS) models \cite{RSOS-1,RSOS-2},
their Hamiltonians have the same structure, see for example
\cite{RSOS-H}. However, in the case of the RSOS theories the Hilbert space is restricted (it consists of certain paths),
while in our case it is simply the tensor product space of the spin chains.

For these models 
we also used the same formulation of our algebraic
methods, as opposed to the ``face weight'' formulation of the Yang-Baxter relations typically used in the RSOS (or IRF)
framework. However, in Section \ref{sec:IRFgates} we showed that the two formulations are indeed identical in these
special cases. This also implies that our
models are solutions to the ``face weight'' formulation of the Yang-Baxter relation.
We believe that the connection between the RSOS models and our new Hamiltonians deserves further study.

The family of the IRF type Hamiltonians accommodates some of the elementary cellular automata that have been
studied recently \cite{rule54-review,prosen-cellaut}. We found that out of the classical cellular automata treated in
\cite{prosen-cellaut} the Rule150 and Rule105 models are Yang-Baxter integrable, and they can be deformed into quantum
cellular automata. We also gave a recipe for computing extensive local charges for
these quantum and classical cellular automata, and we derived the concrete formulas for the first charge of the Rule150
and Rule105 models.
Putting everything together, our construction can be seen as a remarkable 
link between classical and quantum integrable models.

In contrast, we did not find such three site structures for the
famous Rule54 model.
We pointed out that the
construction of \cite{prosen-cellaut} does not yield new conserved charges on top of the known ones, and the transfer
matrices derived there are functionally dependent on the known charges.
Thus the problem of the integrability of the Rule54 model is still open.
A very important piece of the puzzle was presented in \cite{vasseur-rule54}, where a six site interacting Hamiltonian
was constructed, which commutes with the Floquet update rule of the Rule54 model. This suggest that the model could lie
in the family of six site interacting models. Preliminary computations show that this is indeed the case:
we found a new  local extensive charge with interaction range $\ell=10$
using the proper generalization of our methods. 
We will present this result in a future work.

It would be interesting to continue the partial classification of medium range models,  extending our results to more
complicated three site or four
site interacting cases or to higher dimensional local spaces. In both cases a much 
larges parameter space opens up, and a clear physical motivation is needed to formulate the restrictions for the
Hamiltonians. Symmetries can
be chosen as guiding principles, together with special assumptions on the structure of the Hamiltonian. A known four site
interacting model is the folded XXZ model treated in \cite{folded1,folded2,sajat-folded}. This Hamiltonian has a
particular structure: it has two control bits and two action bits, and its algebraic treatment leads to a Yang-Baxter
integrable classical cellular automaton (see \cite{sajat-cellaut} and Section \ref{sec:folded}). It would be interesting to
classify models with a similar structure, potentially leading to new cellular automata with four site update rules.

An other interesting question is whether our constructions exhaust all possibilities for integrable quantum circuits.
The IRF type circuits show very clearly that if the Lax operators have a
special structure, then this allows the construction of special brickwork circuits, which would be meaningless for
other types of Lax operators. Therefore it can not be excluded, that some other sorts of special circuits can be built if we impose some
other special structure on the building blocks. 

In this regard let us return to the so-called box-ball systems mentioned in the Introduction \cite{box-ball,box-ball-review}. These are classical
cellular automata with a less local update procedure, which is performed by acting with a certain transfer matrix which
does not factorize into commuting local unitary operators. Theses systems are special cases of so-called filter automata
\cite{filter1,filter2,filter3,filter4}. The crucial ingredient of such a construction is a Lax operator which becomes
deterministic at some special points, but the regularity condition (that would lead to strictly local update rules)
is not required. Our solutions for the integrable Lax operators could also be used to construct such filter automata.

In this paper we discussed the
physical properties of our models only in passing. We explained that the bond-site transformed XYZ model of Section
\ref{sec:IRFH} describes interacting dynamics of Domain Walls, with or without Domain Wall number conservation. 
And we will publish a paper dealing specifically with the hard rod deformed XXZ model found in Section
\ref{sec:classification}.
We believe that the other new spin chain and quantum gate models also deserve further attention.

Finally let us mention that our methods could be relevant for also the AdS/CFT conjecture. It is known that in the planar limit
the dilatation operator of the gauge theory is essentially an integrable Hamiltonian with long range interaction
\cite{beisert-dilatation0,beisert-dilatation}. The spectrum of this Hamiltonian is 
now understood using the so-called quantum spectral curve method \cite{qsc1,qsc2,gromov-spectral-curve-intro}, but there
is no clear understanding of the actual 
Hamiltonian on the operator level. It is known that it is a long range deformation of an integrable nearest neighbor
chain, but it is not clear how to perform the long range deformation in a finite volume \cite{beisert-long-range-2}. Our
methods could give a recipe for 
this problem: perhaps there is a truncation scheme where we could gradually increase the
interaction range of the chains while still using our present methods at each step. This appears to be a promising
direction for future work.

\begin{acknowledgments}
  We are thankful to Toma\v{z} Prosen for useful discussions, and to 
D\'avid Sz\'asz-Schagrin for computing the level spacing statistics of a spin chain treated in Appendix
\ref{sec:counter}. We are also thankful to Arthur Hutsalyuk and Levente Pristy\'ak for useful comments on the
manuscript. 
\end{acknowledgments}

\appendix

\section{Counter-example to the original conjecture of \cite{integrability-test}}

\label{sec:counter}

Here we discuss a counter-example to the original conjecture of \cite{integrability-test} regarding the integrability of
nearest neighbor spin chains. In \cite{integrability-test}  it was claimed that a sufficient condition of integrability
is the existence of a three-site charge which commutes with the Hamiltonian. However, it was not stressed in
\cite{integrability-test} that the Hamiltonian has to be dynamical.

For example consider the family of models defined by the two charges
\begin{equation}
  \begin{split}
   Q_2&=\sum_j \sigma^z_j \sigma^z_{j+1},\\
   Q_3&=\sum_j  (1-\sigma^z_j \sigma^z_{j+2})(\sigma^x_{j+1}+\kappa\sigma^z_{j+1}).
  \end{split}  
\end{equation}
Here $\kappa\in\valos$ is a coupling constant. Direct computation shows that $[Q_3,Q_2]=0$ for every $\kappa$. The model
is integrable for $\kappa=0$, it is a special point of the bond-site transformed XYZ model considered in Section
\ref{sec:IRFH}. However, a non-zero $\kappa$ introduces a term which our classification found to be non-integrable. To
confirm that the model is indeed non-integrable, we investigated the level spacing distribution for $\kappa=1$ and
confirmed the Wigner-Dyson statistics characteristic for chaotic models. This computation was performed by D\'avid
Sz\'asz-Schagrin and we are thankful to him.

This example shows, that if the nearest neighbor charge $Q_2$ is not dynamical, then the existence of a commuting
charge $Q_3$ is not enough to ensure integrability of the model.

\section{Bond-site transformation}

\label{sec:bond}

A number of models that we encountered in this work together with the folded XXZ model treated in
\cite{folded1,folded2,sajat-folded} allow for an alternative description after performing a bond-site
transformation. This is a non-local transformation, and in the models of interest it leads to Hamiltonians with shorter
interaction range. The bond-site transformation is a special case of the more general Clifford transformations treated recently in
\cite{clifford1}; here we just treat this simple case.

The first observation is that in some models the dynamical Hamiltonian (be it a three-site or four-site operator) commutes
with the non-dynamical charge
\begin{equation}
  Q_2=\sum_{j} \frac{1-\sigma^z_j\sigma^z_{j+1}}{2}.
\end{equation}
Here we chose the conventions such that $Q_2$ has zero eigenvalue on ferromagnetic states in the computational
basis. Then the non-zero contributions to $Q_2$ originate from nearest neighbors where the two spins are
different. Such a situation can be interpreted as a Domain Wall (DW), and then $Q_2$ is seen as the total DW number
which is conserved. It is then natural to expect that the Hamiltonian can be interpreted as an operator that generates
dynamics for the DW's.

This is seen explicitly by performing a bond-site transformation, either in finite volume with open boundary conditions,
or directly in infinite volume. The idea is to put spin-1/2 variables on the bonds between lattice sites, and to perform
a change of basis starting from the original computational basis. For each bond we write down a $\circ$ (up spin) if the two
neighboring spins are identical, and a $\bullet$ (down spin) if they are different. This is a highly non-local transformation,
which can be inverted (up to simple complications at the boundaries). Then we also obtain a new Hamiltonian in the new
basis.

Such a transformation can be
performed for any spin chain with interaction range $\ell$. The new Hamiltonian will be local if the original one
respects spin reflection invariance. Nevertheless, the new interaction will generally have range  $\ell+1$ in the bond
picture. It is important that the bond-site transformation can be performed even if $Q_2$ is not conserved: in this case
Domain Walls can be created or annihilated. The only requirement for the locality of the transformation is the spin-flip
invariance in a given basis.

The utility of the transformation shows itself if the new model  has a smaller interaction range. A
range of $\ell-1$ can be obtained if a further special condition holds: The original
Hamiltonian density should be such that it does not change the spins at the first and the last sites of its
support. Then only the bonds within the support of length $\ell$ are modified, which means that in the bond basis the
interaction range will be $\ell-1$.

A concrete example for the site-bond transformation was presented in \cite{sajat-folded} in the case of the folded XXZ
model. There the first few charges are given by \eqref{foldedQ}. The charge $Q_4$ indeed preserves the first and last
spins, and the domain wall number given by $Q_2$ is also
conserved. After the bond-site transformation the charge $Q_4$ becomes identical to the three-site Hamiltonian
\eqref{Hhrdef2} with $\Delta=0$, which is interpreted as the hard rod deformation of the XX model.

Further examples for the bond-site transformation are presented in Section \eqref{sec:IRFH}. In those cases the original
Hamiltonian is a three-site operator, which preserves the first and the last spins, acting non-trivially on the middle
spin. In certain cases these models can be transformed into a nearest neighbor chain. In these three site interacting
cases the transformation rules for a subset of the allowed operators are found to be
\begin{align}
  \label{bond-site-op}
1\otimes \sigma^x \otimes 1   &\leftrightarrow       \sigma^x \otimes \sigma^x, \\
  -\sigma^z\otimes \sigma^x \otimes \sigma^z    &\leftrightarrow    \sigma^y \otimes \sigma^y,  \\
 \sigma^z\otimes 1\otimes \sigma^z &\leftrightarrow    \sigma^z \otimes \sigma^z, \\
 \sigma^z\otimes \sigma^y \otimes 1 &\leftrightarrow    \sigma^y \otimes \sigma^x, \\
 1 \otimes \sigma^y \otimes \sigma^z &\leftrightarrow    \sigma^x \otimes \sigma^y.
\end{align}
On the l.h.s. above we listed the three site interacting operators allowed by the requirements, which are transformed
into the two site operators on the r.h.s.

\section{Inversion relation for the $R$-matrices}

\label{sec:inv}

Here we show that the regularity property \eqref{eq:regular} of the $R$-matrix and its inversion relation \eqref{Rinv} are 
not independent properties.

Substituting $\lambda_1=\lambda_3$ to the $YB$ equation \eqref{YB} we get 
\begin{multline}
   R_{12}(\lambda_{1},\lambda_2)R_{13}(\lambda_1,\lambda_1)R_{23}(\lambda_2,\lambda_1)=\\
  =R_{23}(\lambda_2,\lambda_1) R_{13}(\lambda_1,\lambda_1) R_{12}(\lambda_{1},\lambda_2).
\end{multline}
Using the regularity property \eqref{eq:regular} we obtain that
\begin{multline}
   R_{12}(\lambda_{1},\lambda_2)R_{21}(\lambda_2,\lambda_1)=\\
  =R_{23}(\lambda_2,\lambda_1) R_{32}(\lambda_{1},\lambda_2).
  \end{multline}
We can see that the left and the right hand sides act trivially on $3$ and $1$ spaces, respectively. Therefore they have
to be equal to an operator $X_2$ acting only on the second space. Writing out the rapidity dependence we get
\begin{align}
  R_{12}(\lambda_{1},\lambda_2)R_{21}(\lambda_2,\lambda_1) &= X_2(\lambda_{1},\lambda_2) \label{eq:unitDer1}\\
  R_{23}(\lambda_2,\lambda_1) R_{32}(\lambda_{1},\lambda_2) &= X_2(\lambda_{1},\lambda_2).
\end{align}
The second equation can be rewritten as
\begin{equation}
 R_{12}(\lambda_2,\lambda_1) R_{21}(\lambda_{1},\lambda_2) = X_1(\lambda_{1},\lambda_2).
\end{equation}
Substituting back to \eqref{eq:unitDer1} we obtain that
\begin{equation}
 X_1(\lambda_2,\lambda_1) = X_2(\lambda_{1},\lambda_2).
\end{equation}
Since the l.h.s. and the r.h.s. act on different spaces the operator $X$ should be proportional to the identity,  i.e.
\begin{equation}
 R_{12}(\lambda,\mu)R_{21}(\mu,\lambda)\sim 1.
\end{equation} 

\section{Factorization property of the $R$-matrix}

\label{sec:factproof}

Here we prove Theorem \ref{thm:fact}.

Substitute $v=0$ to the $RLL$ relation we get
\begin{multline}
R_{A,(12)}(u,0)\La_{A,3}(u)\La_{(12),3}(0)= \\
\La_{(12),3}(0)\La_{A,3}(u) R_{A,(12)}(u,0).
\end{multline}
Now let us use the regularity of the Lax operator to obtain
\begin{equation}
R_{A,(12)}(u,0) \La_{A,3}(u) =
\La_{A,2}(u) \tilde  R_{A,(31)}(u,0).
\end{equation}
After a simple rearrangement we get
\begin{equation}
\La_{A,2}(u)^{-1} R_{A,(12)}(u,0) =
R_{A,(31)}(u,0) \La_{A,3}(u)^{-1}.
\end{equation}
We can see that the l.h.s. and the r.h.s. act trivially on the spaces $3$ and $2$, respectively, therefore they have to
be equal to an operator that acts only on space $1$:
\begin{align}
 \La_{A,2}(u)^{-1} R_{A,(12)}(u,0) &= X_{A,1}(u), \label{eq:factorDeriv1}\\
 R_{A,(31)}(u,0) \La_{A,3}(u)^{-1} &= X_{A,1}(u).
\end{align}
The second equation can be written as
\begin{equation} \label{eq:RwthXL}
 R_{A,(12)}(u,0) = X_{A,2} \La_{A,1}(u).
\end{equation}  
Substituting back to \eqref{eq:factorDeriv1} we obtain that
\begin{equation}
 X_{A,2}(u) \La_{A,1}(u) = \La_{A,2}(u) X_{A,1}(u),
\end{equation} 
therefore
\begin{equation}
 \La_{A,2}(u)^{-1} X_{A,2}(u) = X_{A,1}(u) \La_{A,1}(u)^{-1}.
\end{equation}
Since the l.h.s. and the r.h.s. act trivially on spaces $1$ and $2$ they have to be equal to an operator acting only on
the auxiliary space $A$:
\begin{align}
  \La_{A,2}(u)^{-1} X_{A,2}(u) &= Y_A(u), \\
  X_{A,1}(u) \La_{A,1}(u)^{-1} &= Y_A(u). \label{eq:factorDeriv2}
\end{align}
From the first equation we obtain 
\begin{equation}
 X_{A,1}(u) = \La_{A,1}(u) Y_A(u).
\end{equation}
Substituting back to \eqref{eq:factorDeriv2} we obtain that
\begin{equation}
 [\La_{A,1}(u), Y_A(u)] = 0.
\end{equation}

Substituting back to \eqref{eq:RwthXL} the $R$-matrix reads as
\begin{equation} \label{eq:RfactorDeriv1}
 R_{A,(12)}(u,0) = \La_{A,2}(u) \La_{A,1}(u) Y_A(u).
\end{equation} 
We are almost ready. The only remaining thing is to prove that the operator $Y_A(u)$ has to proportional to the identity.

We also know that the $R$-matrix is regular i.e.
\begin{equation}
 R_{A,(12)}(0,0) = P_{A,(12)},
\end{equation} 
therefore
\begin{equation} \label{eq:RfactorDerivY}
 Y_A(0) \sim 1.
\end{equation}
Let us substitute $\lambda_1=u$, $\lambda_2=v$ and $\lambda_3=0$ to the YB equation.
\begin{multline}
R_{A,B}(u,v) R_{A,(12)}(u,0) R_{B,(12)}(v,0)= \\
R_{B,(12)}(v,0) R_{A,(12)}(u,0) R_{A,B}(u,v).
\end{multline}
Using \eqref{eq:RfactorDeriv1} we obtain that
\begin{multline} \label{eq:RfactorDeriv2}
R_{A,B}(u,v) \La_{A,2}(u) \La_{B,2}(v) \La_{A,1}(u) \La_{B,1}(v) Y_A(u) Y_B(v)= \\
\La_{B,2}(v) \La_{A,2}(u) \La_{B,1}(v) \La_{A,1}(u) Y_A(u) Y_B(v) R_{A,B}(u,v).
\end{multline}
Using the RLL relation on the l.h.s. we obtain that
\begin{multline}
R_{A,B}(u,v) \La_{A,2}(u) \La_{B,2}(v) \La_{A,1}(u) \La_{B,1}(v) Y_A(u) Y_B(v)= \\
\La_{B,2}(v) \La_{A,2}(u) \La_{B,1}(v) \La_{A,1}(u) R_{A,B}(u,v) Y_A(u) Y_B(v).
\end{multline}
Substituting back to \eqref{eq:RfactorDeriv2} we obtain that
\begin{equation}
  [R_{A,B}(u,v), Y_A(u) Y_B(v)] = 0.
\end{equation}
The operator $Y_A(u)$ can only be non-trivial if the $R$-matrix has gauge symmetry (spectral parameter dependent
symmetry).We assumed that such symmetry is excluded therefore 
\begin{equation}
 Y_A(u) = Y_A
\end{equation} 
Using the equation \eqref{eq:RfactorDerivY} we can see that the operator $Y_A$ has to proportional to the identity.


%

\end{document}